\documentclass[onecolumn,sort&compress,numbers]{els-mrw} 

\usepackage{amsmath,amssymb,amsfonts,amsthm,makeidx,graphicx}
\usepackage[labelfont=bf]{caption}
\usepackage{txfonts}
\usepackage{helvet}
\usepackage{authblk}

\newcommand{\be}{\begin{equation}}
\newcommand{\ee}{\end{equation}}
\newcommand{\ba}{\begin{eqnarray}}
\newcommand{\ea}{\end{eqnarray}}
\newcommand{\nl}{\nu_e}
\newcommand{\nm}{\nu_\mu}
\newcommand{\nt}{\nu_\tau}

\newcommand{\anu}{\overline{\nu}_\mu}

\newcommand{\qqbar}{q\overline{q}}
\newcommand{\qbar}{\overline{q}}
\newcommand{\ubar}{\overline{u}}
\newcommand{\dbar}{\overline{d}}
\newcommand{\sbar}{\overline{s}}
\newcommand{\cbar}{\overline{c}}
\newcommand{\bbar}{\overline{b}}

\newcommand{\uubar}{u\overline{u}}
\newcommand{\ddbar}{d\overline{d}}
\newcommand{\ssbar}{s\overline{s}}
\newcommand{\ccbar}{c\overline{c}}
\newcommand{\bbbar}{b\overline{b}}

\usepackage{titlesec}
\renewcommand*\contentsname{}
\begin{document}

\vspace*{2cm}
\begin{center}
{\Huge\bf Key Historical Experiments \\in Hadron Physics}\\
\vspace*{0.7cm}
{\LARGE Claude Amsler}\\
\vspace*{0.7cm}
{\LARGE Marietta Blau Institute for Particle Physics, \\Austrian Academy of Sciences, Vienna, Austria}\\
\end{center}

\tableofcontents

\begin{center}
\end{center}

\begin{abstract}[Abstract]
The experimental observations that led to the development of hadron physics are reviewed with emphasis on the discoveries of mesons and baryons, starting in the 1940s with the pion and kaon which mediate the strong hadronic force.   The evidence for an internal structure consisting of two or three elementary spin $\frac{1}{2}$ particles  will be reviewed.  In 2003 more complex multi-quark hadrons began to emerge. The subsequent developments beyond the early 2000s are covered  in the Review of Particle Physics (Int. J. Mod. Phys. A 41 (2026) 2630011). Given the large number of hadrons, the choice of key experiments is somewhat subjective.  
\end{abstract}

\newpage

\section{Introduction}
Hadrons (from the greek word for `strong') are particles interac\-ting through the strong interaction.  Hadrons are composite structures with finite dimensions, made of {\it partons} (a word introduced in 1969 by R. Feynman), following  results from deep inelastic scattering experiments. There are three types of partons, quarks ($q$), antiquarks ($\bar{q}$) and gluons ($g$).  Quarks have  spin $\frac{1}{2}$ and positive
parities $P$, gluons have spin 1 and negative parities. Hadrons  are bound by gluons and divided into baryons (for `heavy') with half-integer spins, made of three quarks ($qqq$), and mesons (for `medium' heavy) with integer spins, made of quark-antiquark ($q\qbar$) pairs. The electric charges are $\frac{2}{3}$ for the $u, c, t$ quarks and $-\frac{1}{3}$ for the $d, s, b$ quarks. The $u$ and $d$ are `isospin' doublets with isospin $(I, I_3)= (\frac{1}{2}, \pm\frac{1}{2})$, the other quarks have $I=I_3 = 0$. Thus mesons can have isospin $I$ = 0 (isoscalars), 1 (isovectors) or  $\frac{1}{2}$, and the corresponding electric charges are 0,  0 or $\pm 1$, 0 or $\pm 1$, respectively.
The antiquarks have opposite parities, charges and $I_3$. One knows from the processes des\-cribed below (section \ref{sec:DIS}) that quarks and gluons are real entities that, however, remain confined in hadrons.

The parity of a meson is $P$ = ($- 1)^{\ell+1}$, where $\ell$ is the orbital angular momentum of the $q\bar{q}$  pair. The meson spin $J$ is given  by the relation
$|\ell-s| \le J \le |\ell+s|$, where
$s$ = 0 (antiparallel quark spins) or $s$ = 1 (parallel quark spins). The mesons are classified in
$J^{PC}$ multiplets, where $C = (-1)^{\ell+s}$ is the
charge conjugation which flips the charge, and is defined only for  mesons made
of quarks and their own antiquarks (but by convention assigned to all members of the multiplet). The $G$-parity, $G = (-1)^IC$, extends $C$ to the charge triplets (isospin $I = 1$). $P, C$ are conserved quantities in electromagnetic processes, as well as $I$ and $G$ in strong interactions. 
The $\ell = 0$ mesons are the pseudoscalars ($0^{-+}$) and the vectors ($1^{- -}$). 
The orbital excitations $\ell = 1$  are the scalars ($0^{++}$), the axial vectors ($1^{++}$)  and ($1^{+-}$), and the tensors ($2^{++}$).

Baryons have isospin $I=0$, $\frac{1}{2}$ or $\frac{3}{2}$, and positive or negative pari\-ties, depending on the angular momenta between the quarks. The modern theory of strong interaction (Quantum Chromo Dynamics, QCD) predicts other structures to exist and multi-quark hadrons beyond $q\bar{q}$ and $qqq$ have been discovered since the early 2000s (section \ref{sec:exotics}).  

For a comprehensive  information on the hadron spectrum see Ref. \cite{ParticleDataGroup:2026cfk}. 
In the following we shall use the natural units $\hbar=1$ and $c=1$ (hence 1 MeV =$ \frac{1}{197.3}$ fm$^{-1}$).
The hadron masses which are given in the parentheses in MeV are the updated ones listed in Ref.\cite{ParticleDataGroup:2026cfk}.

\section{Discovery of the first light hadrons}
Only two hadrons were known in the early 1940s, the proton and the neutron. The mediator of the interaction between them, the pion $\pi$, was discovered in 1947  in nuclear emulsions exposed to cosmic rays in the Bolivian Andes\footnote{Nuclear emulsions had been exposed before to cosmic rays at high altitude locations (e.g. \cite{HEITLER:1939aa,BOSE:1942aa}). Particle masses were derived from grain density, range and multiple scattering. The `mesotron' tracks reported in ref. \cite{BOSE:1942aa} were probably due to cosmic muons, or possibly to charged pions.} \cite{Lattes:1947mx}. Several events were observed in which a charged comic ray was brought to rest and was decaying into a muon with a constant track length, thus pointing to the two-body decay of a yet unknown particle with mass $m_\pi\simeq 1.30\, m_\mu$  into a muon $\mu$ and an invisible neutrino. The spin of the pion was determined later to be zero by comparing the reaction $pp\to\pi^+d$ to its inverse,  and assuming time reversal invariance \cite{Durbin:1951np}. Although hints of photons converting to $e^+e^-$ pairs from the decay of a neutral meson had been observed earlier in nuclear emulsions, the final evidence for the neutral version of the pion, the $\pi^0$, was obtained at the Berkeley electron synchrotron  \cite{Steinberger:1950equ} by irradiating a target with 330 MeV photons, and observing the accumulation of  events at the minimum opening angle $\alpha$ between the two $\gamma$'s from  $\pi^0\to\gamma\gamma$, which is related to the $\pi^0$ energy $E$ through the relation $\sin(\alpha/2)= m_{\pi^0}/E$. The negative parity of the charged pion was inferred from the observation of the reaction $\pi^-d\to nn $ with stopping  pions in liquid hydrogen,  assuming  the conservation of parity and the antisymmetry of the $nn$  wave function \cite{Panofsky:1950he}. In the rare decay $\pi^0\to (e^+e^-)(e^+e^-)$ the  planes spanned by the two $e^+e^-$ pairs were preferably ortho\-gonal, which implied that the parity of the $\pi^0$ was also negative \cite{Plano:1959zz}. Hence the pion is pseudoscalar  ($J^{PC}=0^{-+}$) since $C(\gamma) = -1$, hence $C(2\gamma) = +1$.

\begin{figure}[htb]
\includegraphics[width=0.48\textwidth]{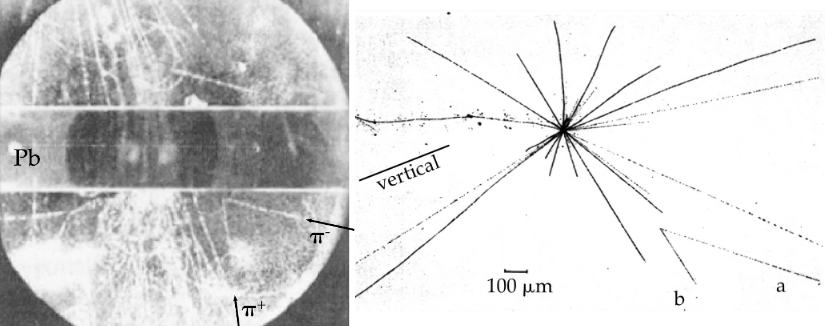}
\centering
\caption[]{Left: Observation of a neutral particle produced by cosmic rays impinging on a lead plate and decaying into two oppositely charged particles, the $K^0\rightarrow\pi^+\pi^-$ (the magnetic field is orthogonal to the picture)  \cite{ROCHESTERDr.:1947aa}. Right:  Observation of a neutral baryon decaying into a meson (track a) and a proton (track b), the $\Lambda\to p\pi^-$. The particle masses were determined from their track lengths  \cite{PhysRev.80.1099}.
\label{K0decay}}
\end{figure}

The kaon $K^+$ had been reported before the pion in cosmic rays with a  mass of 506 $\pm$ 61 MeV  in a cloud chamber imbedded in a magnetic field \cite{leprince1944existence}, and was later confirmed with the observation of the decay  $K^+\to\mu^+\nm $ in nuclear emulsions \cite{doi:10.1080/14786445108561348}. 
The observation of the neutral kaon $K^0$ decaying into two oppositely charged particles was reported  in a cloud chamber exposed to cosmic rays (Fig. \ref{K0decay}, left), but its mass was measured only several years later through  its decay into $\pi^+\pi^-$ \cite{Thompson:1953mk}. The decay of a charged particle with the same mass, but decaying  into three pions, had been reported shortly after the discovery of the pion \cite{Brown:1949mj}. Both particles had spin 0 but opposite parities \cite{Orear:1956aa} and therefore were assumed to be different entities, the $\theta$ decaying to 2$\pi$ and the  $\tau$ decaying to 3$\pi$. The puzzle was solved  in 1957 in favor of one particle only, the kaon, following the non-conservation of parity, observed in nuclear $\beta$-decay \cite{PhysRev.105.1413} and muon decay \cite{PhysRev.105.1415}. Parity violation had been proposed one year before by Lee and Yang \cite{PhysRev.104.254}. 

With the advent of the Cosmotron  at Brookhaven National Laboratory (BNL, 1952 -- 1969) and the Bevatron in Berkeley (1954 -- 2009), accelerators slowly replaced cosmic rays. Cloud chambers and emulsions were supplanted by  bubble chambers (BBC, invented by  D. A. Glaser in 1953) and by electronics detectors. However, the technique of nuclear emulsions (developed in the thirties  by Marietta Blau at the Institute for Radium Research in Vienna) is still in use today in low rate experiments (see {\it e.g.} Ref.\cite{FASER:2024hoe}), due to their unsurpassed (sub-$\mu$m) spatial resolution on ionising tracks \cite{Amsler:2012wn}. 

The four $\Delta(1320)$ baryon resonances ($uuu, uud, udd, ddd$)  which dominate low energy pion-nucleon scattering were discovered at the University of Chicago 450 MeV proton synchrocyclotron \cite{Anderson:1952aa}. The $\pi p$ cross sections was found to rise rapidly towards 200 MeV and the isospin $I=\frac{3}{2}$ was determined from the cross section ratios of 9:2:1 between $\pi^+p\to\pi^+p$, $\pi^-p\to\pi^0n$ and $\pi^-p\to\pi^-p$.

The $\Lambda(1116)$ baryon decaying to $\pi^-p$ was discovered in a nuclear  emulsion during a balloon flight (Fig. \ref{K0decay}, right) \cite{PhysRev.80.1099}.   In the reaction  $\pi^-p\to K^0 \Lambda$, studied at the Cosmotron \cite{PhysRev.98.121}  in a pion beam,  the magnitude of the cross section  was typical of the  strong interaction ($\sim$1~mb at 1 GeV/c).  On the other hand, the decay lengths of the $K^0$ and $\Lambda$ were in the range of a few cm, typical  
of the weak interaction. The large difference in interaction strengths bet\-ween 
production and decay was taken into account by Gell-Mann, Nakano and Nishijima,  by introducing the additive quantum number, `strangeness' $S$.
The $K^0$ was assigned $S$ = +1 and the $\Lambda$  strangeness $S$ = --1 (other hadrons, such as  the $\pi$ or the $p$, had 
$S$ = 0). The total strangeness is conserved in production (strong interactions), but changes by one unit in $\Lambda$ and  $K^0$ decays (weak interactions). We now know that strangeness is due to the $\bar{s}$ antiquark in the $K^0$ and the $s$ quark in the $\Lambda$ with strangeness $S=+1$ and $-1$, respectively. Hyperons are baryons with $S\neq 0$: The $\Lambda$ (1116) was the first one established, followed by  the $\Sigma(1189)^+$ , $\Sigma(1197)^-$  and many more \cite{ParticleDataGroup:2026cfk}. The $\Sigma(1193)^0\to\Lambda\gamma$ was observed in a cloud chamber at the Cosmotron \cite{PhysRev.98.1407}.

The combined operation $CP$ of charge-conjugation and parity  transforms a $K^0$ into its antiparticle, the $\bar{K}^0$. Since $CP$ is (in good approximation) a conserved symmetry, $K^0$ and $\bar{K}^0$ cannot be the observed states, but must be combinations of the $CP$ eigenstates, the $K_1$ and $K_2$, which (again in good approximation) are the $K_S$ and $K_L$, decaying into  $2\pi$ and $3\pi$, respectively, the latter with a very long lifetime due to the limited decay phase space. The lifetime of the $K_S$ (Fig. \ref{K0decay}  with a track length in the cm range) has a  mean life of $\tau_S = 90$ ps. The $K_L$ was observed at the Cosmotron and its mean life ($\tau_L =  51$ ns) was measured at the Princeton-Pennsylvania accelerator with an electronic detector  on a sliding trolley \cite{Vosburgh:1972zqy}.

\begin{figure}[htb]
\includegraphics[width=0.48\textwidth]{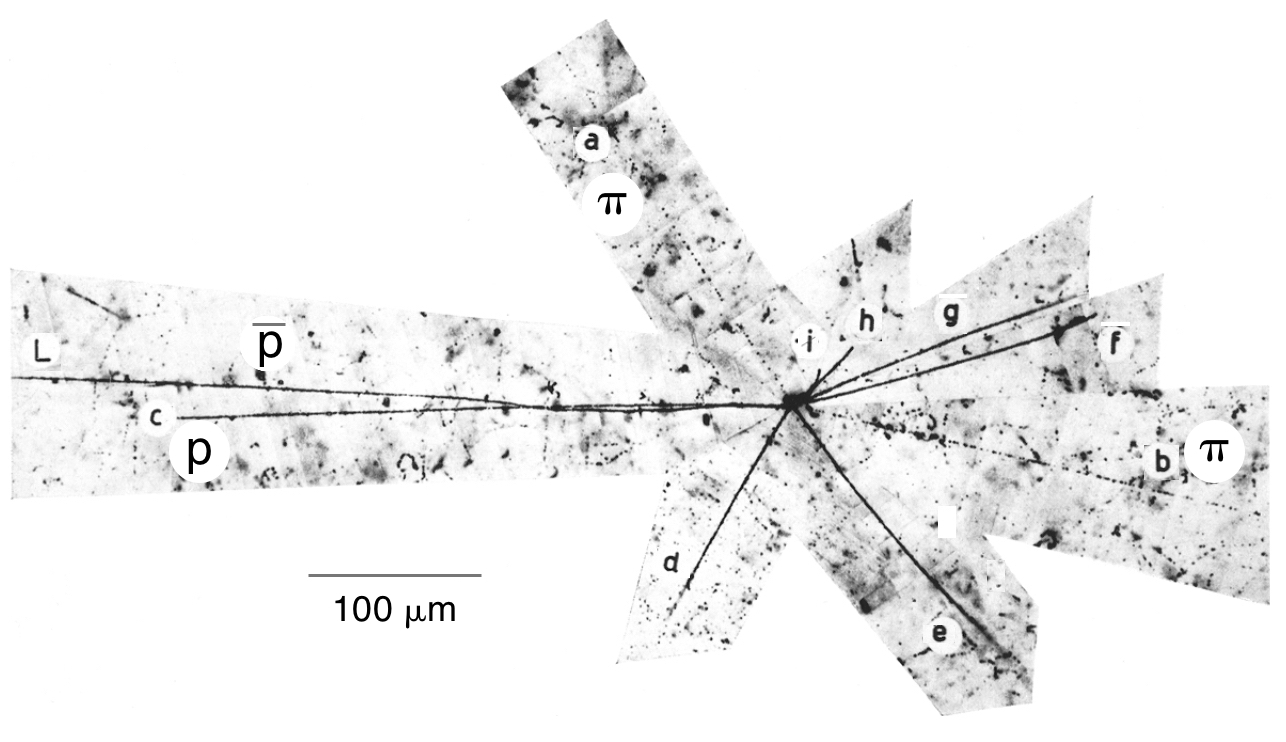}
\centering
\caption[]{The `star' produced by the annihilation of a stopping $\bar{p}$  in a nuclear emulsion (left track L). Tracks $a, b$ are pions, $c$ a proton, the remaining tracks are protons or $\alpha$'s  \cite{Chamberlain:1956aa}.
\label{pbaremul}}
\end{figure}

Following the discovery of the charge conjugated partner of the electron, the positron $e^+$, it was expected that the spin $\frac{1}{2}$ proton  would also have an antipartner. This  was not  obvious, the $g$-factor of the proton being  much larger than the expected $g$ = 2. To produce an antiproton ($\bar{p}$)  by shooting high energy protons on protons, baryon number conservation demands the creation of a $\bar{p}p$ pair ($pp\to ppp\bar{p}$), thus requiring incident protons with kinetic energies of at least $6\,m_p$ = 5.63 GeV. Searching for the antiproton was the main motivation for the Berkeley 6.3 GeV Bevatron proton accelerator. A negatively charged secondary beam from an internal target was commissioned  in 1955.  Antiprotons had to be sought like needles in a haystack (1 $\bar{p}$ in a beam of $\sim5\times$10$^4$ $\pi^-$). Since antiprotons would be much slower than pions, the expected $\bar{p}$ velocity was selected by recording the time of flight between two scintillation counters and, as a further suppression of pions, by \v{C}erenkov counters. This led to the observation of a few dozen negatively charged particles with masses within 5\% of the proton mass  \cite{Chamberlain:1955ns}. No such particle living  long enough to reach the last counter was known. The discovery was  strengthened by the observation of $\bar{p}$ annihilation in nuclear emulsions (Fig. \ref{pbaremul}). The mass of the incident projectile  (0.99 $\pm$ 0.03 $m_p$) was derived by various methods, from grain density ($dE/dx$), multiple scattering ($\beta p$) and range.

We now know that on average 3 charged pions and 2$\pi^0$ are produced in $\bar{p}p$ annihilation at rest. The yield of $K$ and $\eta$ mesons  is much smaller (each about  7 \% of all annihilation). For recent data on $\bar{p}p$ annihilation at rest on various nuclei at the CERN AD/ELENA facility see Ref.\cite{Amsler:2024aa}.

\subsection{Mass measurements}
The  proton mass was initially determined by mass spectrometry, but the best values are obtained from the rotation (cyclotron) frequency in magnetic traps \cite{ParticleDataGroup:2026cfk}.  Resonance masses are determined from the momenta and energies of the decay products. The masses of  long-lived hadrons, such as the $\Sigma^+$, were determined in nuclear emulsions by measuring the range of the daughter proton from stopping $\Sigma^+$ decaying to $p\pi^0$, or in $K^-p\to\Sigma^+\pi^-$ \cite{Barkas:1963zz}. An alternative method applicable to negatively charged particles is to capture the hadron in a nuclear target and to detect  the X-rays emitted by its cascade down the atomic levels. The energy of the X-ray emitted from  level $n$ to  level $n-1$ is 
\begin{equation}
E_\gamma = \left(\frac{m_hM}{m_h+M}\right)\frac{\alpha^2Z^2}{2}\left[\frac{1}{(n-1)^2}-\frac{1}{n^2}\right],
\end{equation}
where $\alpha$ = 1/197.3  is the fine structure constant, $m_h$ the hadron mass and $M$ that of the nucleus. The $\Sigma^-$ and $K^-$ masses were measured this way at BNL with Pb and W targets \cite{Gall:1988ei} . Plans are underway by the SIDDHARTA-2 collaboration at  DA$\Phi$NE to improve on the $K^-$ mass uncertainty with kaons from  $\phi\to K^+K^-$ \cite{Bosnar:2024ibq}. 

\section{Three quarks for Muster Mark!}
Inspired by this line of verse from James Joyce$'$s novel Finne\-gans Wake (1939), Gell-Mann introduced in 1963  the word `quark' and the SU(3) symmetry, together with  Ne$'$eman, Petermann and Zweig \cite{Gell-Mann:1964ewy}, first as a bookkeeping tool to classify the increasing number of observed hadrons discovered at the Bevatron and at two new proton synchrotrons, the AGS (33 GeV) at BNL  and the PS (24 GeV) at CERN.  As we shall see in section \ref{sec:DIS}, quarks are nevertheless  real entities that, however, remain confined in hadrons. 

\subsection{The ground state hadrons}
We now recall the discoveries of the light ground state mesons (apart from the $\pi$ and $K$), {\it i.e.}  those without excitation of the $q\bar{q}$ pair.
\begin{itemize}
\item
$\omega$(782): The  $J^{PC}$=$1^{--}$ isoscalar meson (parallel quark spins) was discovered at the Bevatron with 1.6 GeV/c  $\bar{p}$ annihilating at rest into  $2\pi^+2\pi^-\pi^0$  in a hydrogen BBC  \cite{Maglich:1961rtx}. Fig. \ref{omegam} (top) shows the $\pi^+\pi^-\pi^0$ invariant mass distribution  revealing the narrow $\omega$. The absence of a signal in $\pi^\pm\pi^\pm\pi^\mp$ (charge $Q=1$) and  $\pi^\pm\pi^\pm\pi^0$ ($Q=2$) proved that the $\omega$  had $I$ = 0.  Conservation of $G = -1$ in the $3\pi$ decay implied that $C=-1$. There was reasonable evidence for $J^P$=$1^-$ from the energy sharing between the three pions (Dalitz plot), but the final confirmation came with more data \cite{Stevenson:1962zz}. 

\begin{figure}[htb]
\includegraphics[width=0.48\textwidth]{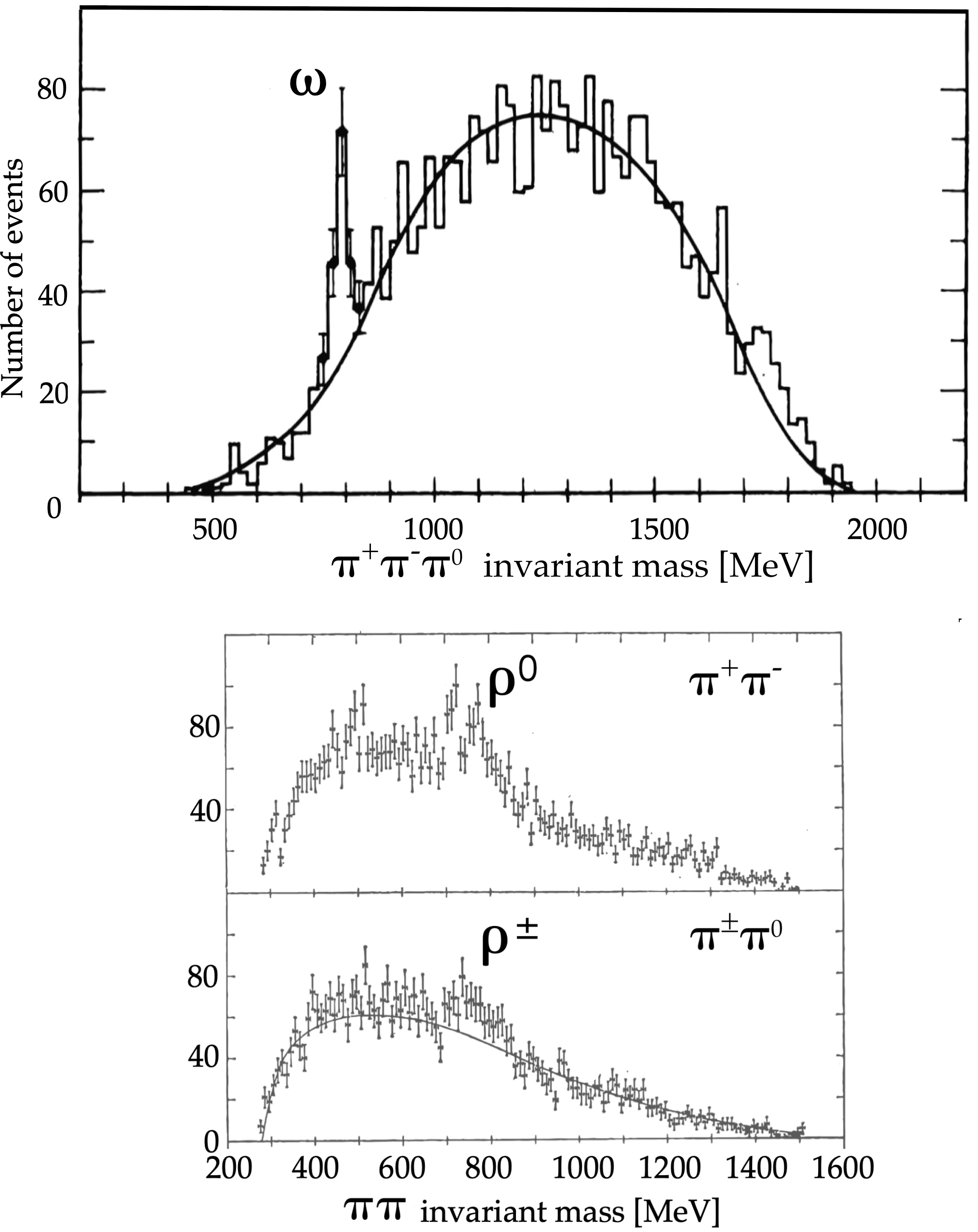}
\centering
\caption[]{Top:  $\omega\to\pi^+\pi^-\pi^0$  in $\bar{p}p$ annihilation into $2\pi^+2\pi^-\pi^0$ (4 entries/event, adapted from \cite{Maglich:1961rtx}). Bottom: Observation of  $\rho^0\to\pi^+\pi^-$ (top) and $\pi^\pm\pi^0$ (bottom) in $\bar{p}p$ annihilation into $2\pi^+2\pi^-n\pi^0$  \cite{Button:1962bnf}.
\label{omegam}}
\end{figure}

\item
$\rho$(770): The production cross section of isovector $\pi\pi$-pairs in inelastic $\pi^-p$ scattering appeared to be resonating in the 780 MeV region \cite{Anderson:1961zz}. Fig. \ref{omegam} (bottom) shows the 2$\pi$-invariant mass distribution in $\bar{p}p\to 2\pi^+2\pi^-n\pi^0$ with 1.6 GeV/c $\bar{p}$ in a BBC at the Bevatron.  A 110 MeV broad peak appears around 770 MeV, the  $\rho$ meson,  in  $\pi^\pm\pi^0$ (hence $I = 1$) and $\pi^+\pi^-$,  hence $C = -1$ since $G = +1$ \cite{Button:1962bnf} \footnote{The hint of a double structure in the $\pi^+\pi^-$ mode (Fig. \ref{omegam}, bottom)  could also be the first observation of $\rho-\omega$ interference, where the $\omega$ decay  into $\pi^+\pi^-$ is  electromagnetic and  isospin violating, a process  confirmed many years later  \cite{Allison:1970uu}.}.
\item
$\phi$(1020): The reactions $K^-p\to \Lambda K^0\bar{K}^0 $ and $\Lambda K^+K^-$ were studied in a BBC exposed to 1.2 and 2.5 GeV/c kaons at the AGS \cite{Connolly:1963pb}. A 10 MeV narrow  signal   with 42 events was observed in the $K^0\bar{K}^0 $ and $K^+K^-$ mass spectra at 1019 MeV, the $\phi$. The $K^0\bar{K}^0 $ was consistent with $K_1K_2$, which can be shown to have $C=-1$.  This also meant that $P=-1$, implying odd spin and presumably $J=1$
in view of the limited decay phase space. Since a signal hint appeared  in $\pi^+\pi^-\pi^0$ associated with a $\Lambda$, but not in $\pi^+\pi^-$, the conclusion was that $G=-1$ and hence $I = 0$. 
\item
$K^*$(892): The spin 1 version of the  kaon (parallel quark spins) was observed at the Bevatron in the reaction $K^-p\to K^{*-} p$ with $K^{*-} \to \bar{K}^0\pi^-$ \cite{Alston:1961nx} with a resonating $K\pi$ $P$-wave, hence $J^P = 1^-$ \cite{Wojcicki:1964zz}.
\item
$\eta$(548): A  BBC exposure to 1.2 GeV/c pions at the Bevatron  revealed in the reaction $\pi^+d\to pp\,(\pi^+\pi^-\pi^0)$ besides the $\omega$  a further 25 MeV signal around 548 MeV. This was the first observation of the isoscalar $\eta\to\pi^+\pi^-\pi^0$ \cite{Pevsner:1961pa}, with no signal appearing in $\pi^+d\to pn\,(\pi^+\pi^+\pi^-)$. Its $\gamma\gamma$ decay was observed at the Cosmotron with 1.1 GeV/c pions in $\pi^-p\to n \eta$ \cite{PhysRevLett.9.127}.  The mass of $\simeq$ 545 MeV was derived from the minimum opening angle between the $2\gamma$. A spin 1 meson cannot decay into 2$\gamma$ (Fermi-Yang theorem) and the absence of $\pi\pi$ decay mode excludes $J^P$ = $0^+$, $2^+$... thus leaving as most likely $0^-$ (its 3$\pi$ decay is isospin and $G$-violating due to $m_u\neq m_d$).
\item
$\eta'$(958): Another narrow isoscalar state was discovered near 960 MeV at the Bevatron \cite{Goldberg:1964zza} and the AGS \cite{Kalbfleisch:1964zz} from BBC studies of $K^-p\to\Lambda $+ multipions. The dominant decay mode was $\eta'\to\eta\,(\to\pi^+\pi^-\pi^0)\,\pi^+\pi^-$ \cite{Goldberg:1964zza}. The spin-parity  was resolved later unambiguously in favor of ($0^-$) by a BBC exposure at the CERN-PS  from the the $\eta'$ decay angular distribution in $K^-p\to \Lambda \eta'$ \cite{Amsterdam-CERN-Nijmegen-Oxford:1977xfz}.
\end{itemize}
The ground state mesons made of the $u, d, s$ light quarks are summarized in Table \ref{tab:SU3}.   According to SU(3), the $\eta$, $\eta'$ and $\phi$, $\omega$ pairs are described by  rotations of the wave functions $\psi_1= (u\bar{u} + d\bar{d} + s\bar{s})/\sqrt{3}$ and $\psi_8 =  (u\bar{u} + d\bar{d} - 2 s\bar{s})/\sqrt{6}$. The rotation angles are determined {\it e.g.} from the meson masses  as $\theta_0\sim -18^\circ$ and $\theta_1\sim 38^\circ$, respectively  \cite{ParticleDataGroup:2026cfk}. It easy to show that $\phi$ is almost purely $s\bar{s}$ and $\omega$ mostly $d\bar{d} + u\bar{u}$.  About 100 orbital and radial excitations of the ground state mesons have been discovered  \cite{ParticleDataGroup:2026cfk}.  

\begin{table}[htb]
\begin{center}
\caption[]{The ground state light mesons (following SU(3) symmetry). The SU(3) wave function of the $\eta'$ is orthogonal to that of the $\eta$, the wave function of the $\omega$ is orthogonal to that of the $\phi$.}
\footnotesize
\begin{tabular}{l l l l l}
\hline
& $I^{[G]}(J^{P[C]})$ & &$I^{[G]}(J^{P[C]})$ & SU(3) wave function\\
\hline
$\pi^+$, $\pi^-$ & $0^-(0^-)$ & $\rho^+$, $\rho^-$ & $1^+(1^-)$ & $u\bar{d}$, $\bar{u}d$ \\
$\pi^0$ & $0^-(0^{-+})$ & $\rho^0$ & $1^+(1^{--})$ & $(d\bar{d} - u\bar{u}) /\sqrt{2}$\\
$K^+$, $K^-$ & $\frac{1}{2}(0^{-})$ & $K^{*+}, K^{*-}$ & $\frac{1}{2}(1^{-})$ & $\bar{s}u$, $s\bar{u}$\\
$K^0$, $\bar{K}^0$ & $\frac{1}{2}(0^{-})$ & $K^{*0}$, $\bar{K}^{*0}$ & $\frac{1}{2}(1^{-})$ & $\bar{s}d$, $s\bar{d}$\\
$\eta$, $\eta'$ & $0^+(0^{-+})$  &   &  & $\eta = \psi_8\cos\theta_0 - \psi_1 \sin\theta_0$\\ 
& & $\phi$, $\omega$ & $0^-(1^{--})$   & $\phi = \psi_8\cos\theta_1 - \psi_1 \sin\theta_1$\\
\hline
\end{tabular}
\label{tab:SU3}
\end{center}
\end{table}

\begin{figure}[htb]
\includegraphics[width=0.40\textwidth]{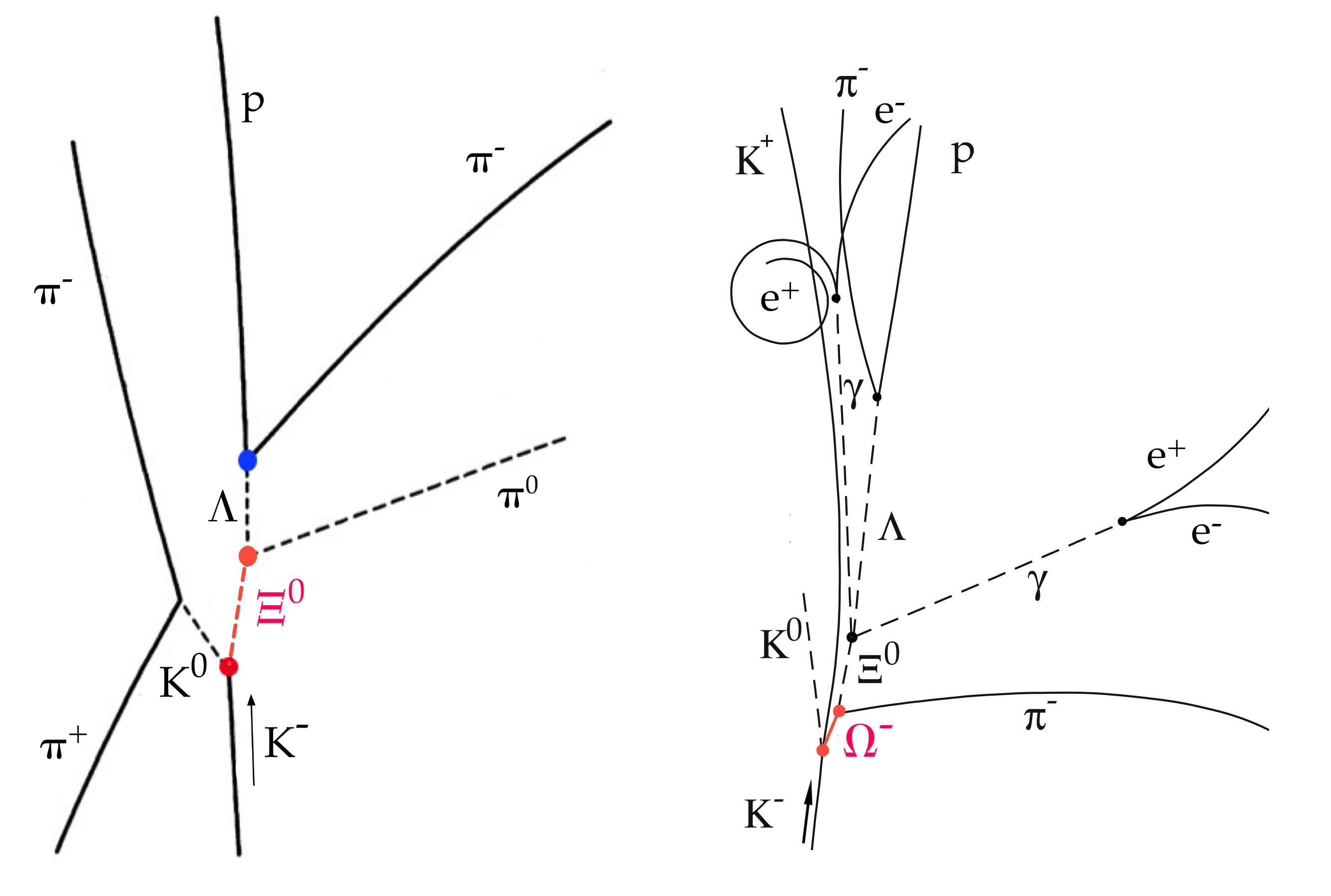}
\centering
\caption[]{Left: production and decay of the $\Xi^0$ in a BBC \cite{Alvarez:1959zz}. Invi\-sible neutral particles are shown by dashed lines. The line joining the $\Xi^0$ production point and the $\Lambda$ decay vertex does not lie in the plane spanned by the $\pi^-$ and $p$ tracks,  thus an invisible particle (the $\pi^0$) must have been emitted. Right: BBC photograph showing the production and decay of the $\Omega^-$  
\cite{Barnes:1964pd}.
\label{XiOm}}
\end{figure}

We have already mentioned the $\Lambda$, $\Sigma^\pm$ and $\Sigma^0$ hyperons with strangeness $S=-1$. Fig. \ref{XiOm} (left) shows the  $\Xi(1315)^0$, a strangeness  $S=-2$ ($ssu$) hyperon decaying to $\Lambda\pi^0$, as observed at the  Bevatron with 1.15 GeV/c  kaons, in   $K^-p\to K^0\Xi^0$ \cite{Alvarez:1959zz}. Fig. \ref{XiOm} (right) shows the discovery  of the   $\Omega(1672 )^-$ hyperon, $S=-3$ ($sss$)   \cite{Barnes:1964pd}. It was produced with 5 GeV/c kaons from the AGS in $K^-p\to \Omega^-K^+K^0$, and decayed into  $\Xi^0\pi^-$, in turn followed by  $\Xi^0\to\Lambda\pi^0$. The observation of hyperons with  3 strange quarks was hailed as a triumph of the quark model. 

\section{Measurement of the proton size}
\label{Radius}
In the celebrated experiment of  Geiger and Marsden (Rutherford$'$s collaborators at the University of Manchester) \cite{Geiger1913LXITL} $\alpha$ particles of 4.78 MeV from a radium source were scattered off a thin gold foil and detected by a ZnS scintillator as a function of scattering angle $\theta$. Surprising was the observation of backward scattered $\alpha$ particles which, according to Rutherford, was like `shooting a cannon ball at a piece of tissue paper and have the ball bouncing back at you'. This was the discovery of the hard core inside the gold atom, the atomic nucleus.

Gold nuclei and $\alpha$ particles being spinless, let us consider electromagnetic elastic scattering of two spin 0 particles, first both structureless (point-like) and with unit electric charges. The process is mediated by the exchange of a photon between projectile and target (Fig. \ref{deepin}a below). The measured differential cross section in the laboratory reference frame  is given by the Rutherford scattering formula
\be
\left(\frac{d\sigma}{d\Omega}\right)= 
\frac{4\alpha^2m^2}{q^4},
\label{Ruther}
\ee
(neglecting the target recoil energy) where  $m$ is the projectile mass and $q=2p\sin\frac{\theta}{2}$, the momentum transferred by the exchanged photon, with $p$ the incident momentum.  For nuclei of finite dimensions eq. (\ref{Ruther}) fails when the momentum is high enough for the projectile to enter the target volume. In the Geiger-Marsden experiment the distance of  closest approach was 27 fm (for $\theta = 180^\circ$)  compared to a gold radius  of 7 fm. Hence higher incident momenta are needed to sample  the nuclear structure  and  point-like projectiles such as  electrons are preferred to avoid having to disentangle projectile and target.

Next, let us consider elastic electron-proton scattering with a spin $\frac{1}{2}$ electron, but still ignoring the proton spin.  To measure the proton size one needs a De Broglie wavelength $2\pi/p$ comparable to the size of the proton, hence electrons of a few 100 MeV/c. The differential cross section in the laboratory is obtained from eq. (\ref{Ruther}) by (i) replacing $m$ by the energy $E\simeq p$ of the electron, (ii) multiplying by $\cos^2\frac{\theta}{2}$ to preserve the electron spin helicity (see {\it e.g.} Ref.\cite{10.1088/978-0-7503-1140-3}) and  (iii) taking into account the proton recoil energy by multiplying with the factor
\be
\frac{E'}{E} = \frac{M}{M+2E\sin^2\frac{\theta}{2}},
\label{recoil}
\ee 
where $E'$ is the energy of the scattered electron and $M$ the proton mass. The Mott formula is given by
\be
\left(\frac{d\sigma}{d\Omega}\right)_M = 
\frac{4\alpha^2E'^2}{Q^4}\left(\frac{E'}{E}\right)\cos^2\frac{\theta}{2}.
\label{Mott}
\ee 
We have replaced $q^4$ by $Q^4 = [2M(E-E')]^2 = [4EE'\sin^2\frac{\theta}{2}]^2 = q^4E'^2/E^2$, where  $Q$ is the
 4-momentum transfer.

There are two additional complications: Firstly, the proton possesses a magnetic moment with $g$-factor $g_p = 5.58$ with which the magnetic moment of the electron interacts. Secondly, there are two form factors which take into account the charge  and magnetic distribution $\rho(\vec{r}\,)$ inside the proton, $G_E(Q^2)$ and $G_M(Q^2)$. (The form factors  arise from the diffraction of the incident plane wave by the proton, akin to Fraunhofer diffraction of light traversing a pinhole.) The electric or magnetic charge distributions are obtained from the measured form factors,  which are the Fourier transforms $\int\rho(\vec{r}\,)\,\exp(i\vec{q}\cdot\vec{r}\,)\,d^3\vec{r}$. 
The differential cross section for elastic scattering $ep\to ep$ is given by the Rosenbluth formula

\be
\left(\frac{d\sigma}{d\Omega}\right)_R = 
\frac{4\alpha^2E'^2}{Q^4}\frac{E'}{E}\left[\frac{G_E^2(Q^2)+\tau 
G_M^2(Q^2)}{1+\tau}\cos^2\frac{\theta}{2} + 2\tau 
G_M^2(Q^2)\sin^2\frac{\theta}{2}\right] \ ,
\label{Rosen}
\ee
with $\tau = Q^2/4M^2$. For forward scattering $G_E(0)$ = 1, $G_M(0) = g_p/2$. For $g_p$=2 (Dirac particle)  one retrieves eq.  (\ref{Mott}) at small scattering angles.

\begin{figure}[htb]
\includegraphics[width=0.48\textwidth]{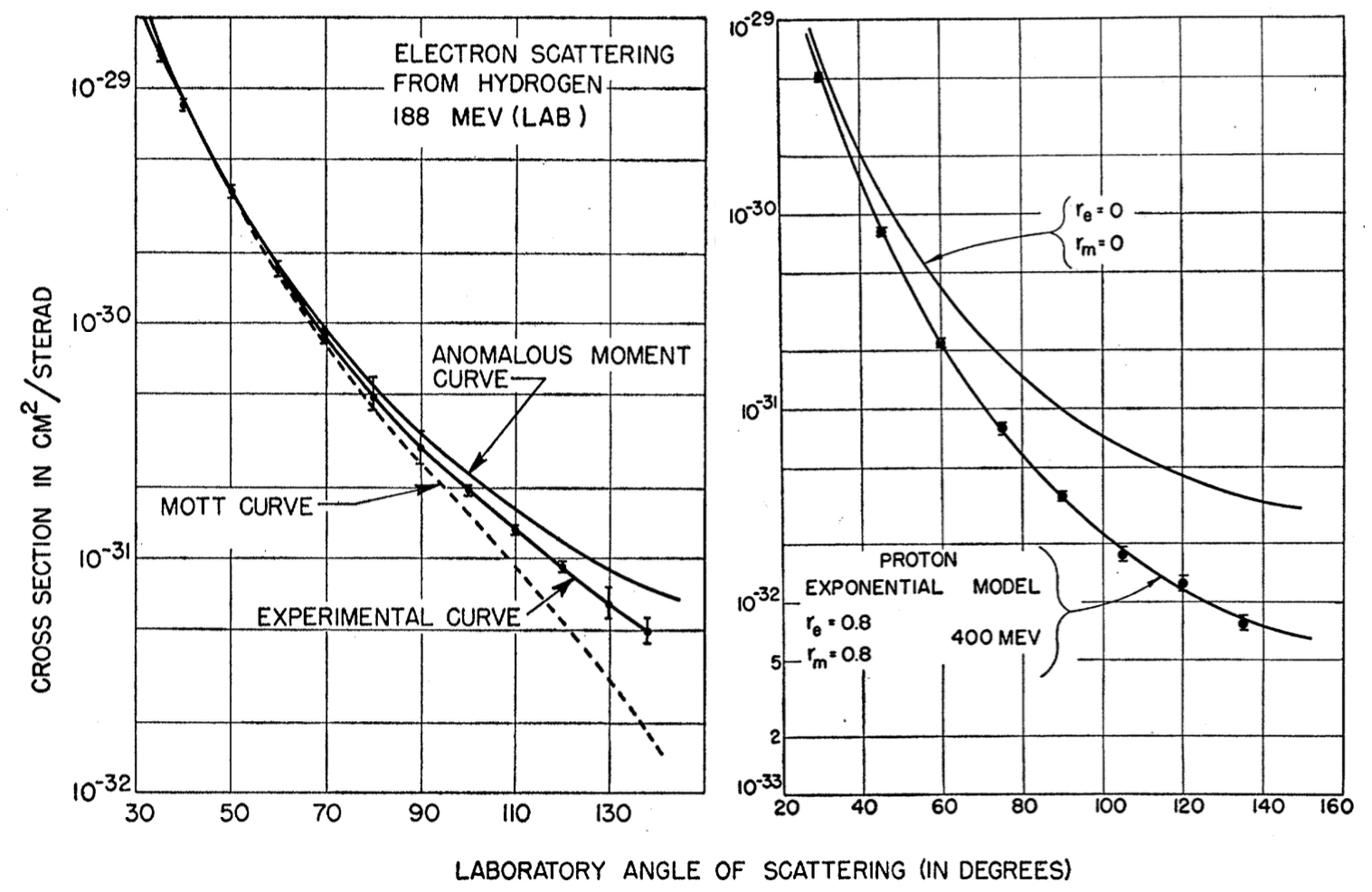}
\centering
\caption[]{Left: $ep$ elastic differential cross section at 188 MeV, leading to $r_e =  r_m$ = 0.70 $\pm$ 0.24 fm  \cite{PhysRev.98.217}. Right: Angular distribution at 400 MeV. The data are fitted with exponential charge and magnetic distributions, leading to rms radii $r_e$ and $r_m$ of 0.80 $\pm$ 0.05 fm \cite{RevModPhys.28.214}.
\label{Hofst}}
\end{figure}

The first measurements  were performed at the Stanford Linear Accelerator (SLAC) by R. Hofstadter \cite{PhysRev.98.217}. The electrons were directed into a  gaseous hydrogen chamber and the ones scattered under the angle $\theta$ momentum analyzed by a magnetic spectrometer installed on a movable platform. Although not independent, see eq. (\ref{recoil}), both $\theta$ and $E'$ were measured  to remove background from inelastic scattering. Fig. \ref{Hofst} (left) shows the cross section measured with 188 MeV electrons. The data deviate from the Mott prediction and also from the point-like charge and magnetic moment distributions (the anomalous moment curve refers to the prediction from eq. (\ref{Rosen}) with $G_E \equiv G_M \equiv  1$). The `dipole' form factor $F(q)=1/(1+a^2q^2)^2$ is often used to fit the proton data (Fig. \ref{Hofst}, right). It corresponds to a  charge and magnetic distribution $\sim\exp(-r/a)$ with the r.m.s proton radius given by $\sqrt{\langle r^2\rangle} = \sqrt{12}a$.

The radii of the charged pion and kaon ($\simeq$ 0.66 fm) were determined at the CERN SPS by scattering $\sim$300 GeV pions and kaons off electrons in a liquid hydrogen target \cite{AMENDOLIA1984116,AMENDOLIA1986435}.

\section{Experimental evidence for spin $\frac{1}{2}$ quarks}
\subsection{Deep inelastic electron-proton scattering}
\label{sec:DIS}
\begin{figure}[htb]
\includegraphics[width=0.40\textwidth]{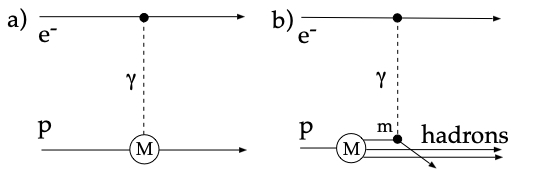}
\centering
\caption[]{Elastic (a) and deep inelastic (b) electron-proton scattering in which the exchanged photon strikes a quark of mass $m$.
\label{deepin}}
\end{figure}

In two-body elastic scattering only one variable needs to be speci\-fied in the final state, hence $E'$ and $\theta$ are not independent, but related through eq. (\ref{recoil}).  Introducing the variable $\nu=  E - E'$ leads to an important relation between $\nu$ and  $Q^2$: $\nu = \frac{Q^2}{2M}$.
With increasing incident energy and momentum transfer, excited proton states such as the $\Delta(1232)$, are produced. With yet higher
momentum transfer, and correspondingly smaller  wavelength of the exchanged photon, one enters the regime of deep inelastic scattering, in which the hypothetical constituents (the  quarks) are resolved. The exchanged photon strikes a quark of mass $m$, which we express as a fraction of the proton mass $M$: $m=xM$. The proton then breaks apart into partons which then `fragment' into hadrons (Fig. \ref{deepin}b). 

Two variables can now be chosen independently, such as $\theta$ and $E'$, or $\nu$ and $Q^2$. Let us  assume that the electron scatters elastically off a  `point-like' spin $\frac{1}{2}$ quark with  mass $m = x M$ inside the proton, and use  eq. (\ref{Rosen}) without the form factors.  The second derivative of the cross section is given by\footnote{Replacing $m$ by $M$ and integrating over $E'$ reproduces the point-like Rosenbluth formula, for a derivation see {\it e.g.} \cite{HalzenMartin,10.1088/978-0-7503-1140-3}.}

\ba
\frac{d^2\sigma}{d\Omega dE'} & = &
\frac{4\alpha^2E'^2}{Q^4}\left[\cos^2\frac{\theta}{2} +  \frac{Q^2}{2m^2}
\sin^2\frac{\theta}{2}\right]\delta\left(\frac{Q^2}{2m}-\nu \right)  \nonumber  \\ 
& = & \frac{4\alpha^2E'^2}{Q^4}F_2(x)\left[\frac{1}{\nu}\cos^2\frac{\theta}{2} +  \frac{1}{xM}\sin^2\frac{\theta}{2}\right],
\label{eq:deepin}
\ea

\noindent
with $x=\frac{Q^2}{2M\nu}$ and the `structure function' $F_2(x) = x\,\delta\left(x - \frac{Q^2}{2M\nu}\right)$, where $\delta$ is the Dirac delta-function. 
$F_2(x)$ needs  to be folded with the probability distribution of the quark mass. Let  $q_i(x)$ describe the $x$-distribution of the quarks with flavor $i$ and  charge $eQ_i$. Summing over all quark flavors and folding with the $\delta$-function leads to
\be
F_2(x) = \int_0^1 \sum_i q_i(x')x'\delta\left(x'-\frac{Q^2}{2M\nu}\right) Q_i^2 dx' = \sum_i q_i(x)x Q_i^2.
\label{eq:F2x}
\ee

The cross section for inelastic scattering on a  proton with diffuse structure (no internal constituents) should include two form factors, functions  of the two independent variables $Q^2$ and $\nu$.  On the contrary,  eq. (\ref{eq:deepin})  for scattering off point like constituents contains only one structure function which depends on one dimensionless variable only, the Bjorken `scaling' variable $x$.

\begin{figure}[htb]
\includegraphics[width=0.30\textwidth]{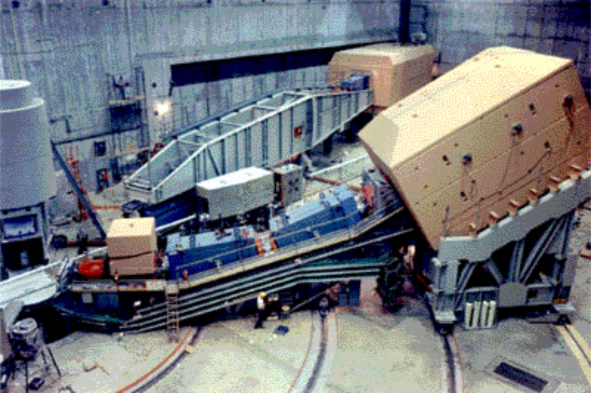}
\centering
\caption[]{Photograph of the experimental hall showing the 8 and 20 GeV spectrometers (Image credit: SLAC National Accelerator Laboratory).
\label{SLAC}}
\end{figure}

Deep inelastic electron-proton scattering was measured between 1967 and 1973 by an MIT-SLAC collaboration at SLAC  with magnetic spectrometers \cite{annurev:/content/journals/10.1146/annurev.ns.22.120172.001223}.  Electrons up to 20 GeV were scattered in a liquid hydrogen target and were momentum analyzed by the magnetic spectrometers (Fig. \ref{SLAC}). Particle identification to distinguish electrons from pions was achieved with gas \v{C}erenkov counters and total absorption counters measuring electromagnetic showers. 

\begin{figure}[htb]
\includegraphics[width=0.43\textwidth]{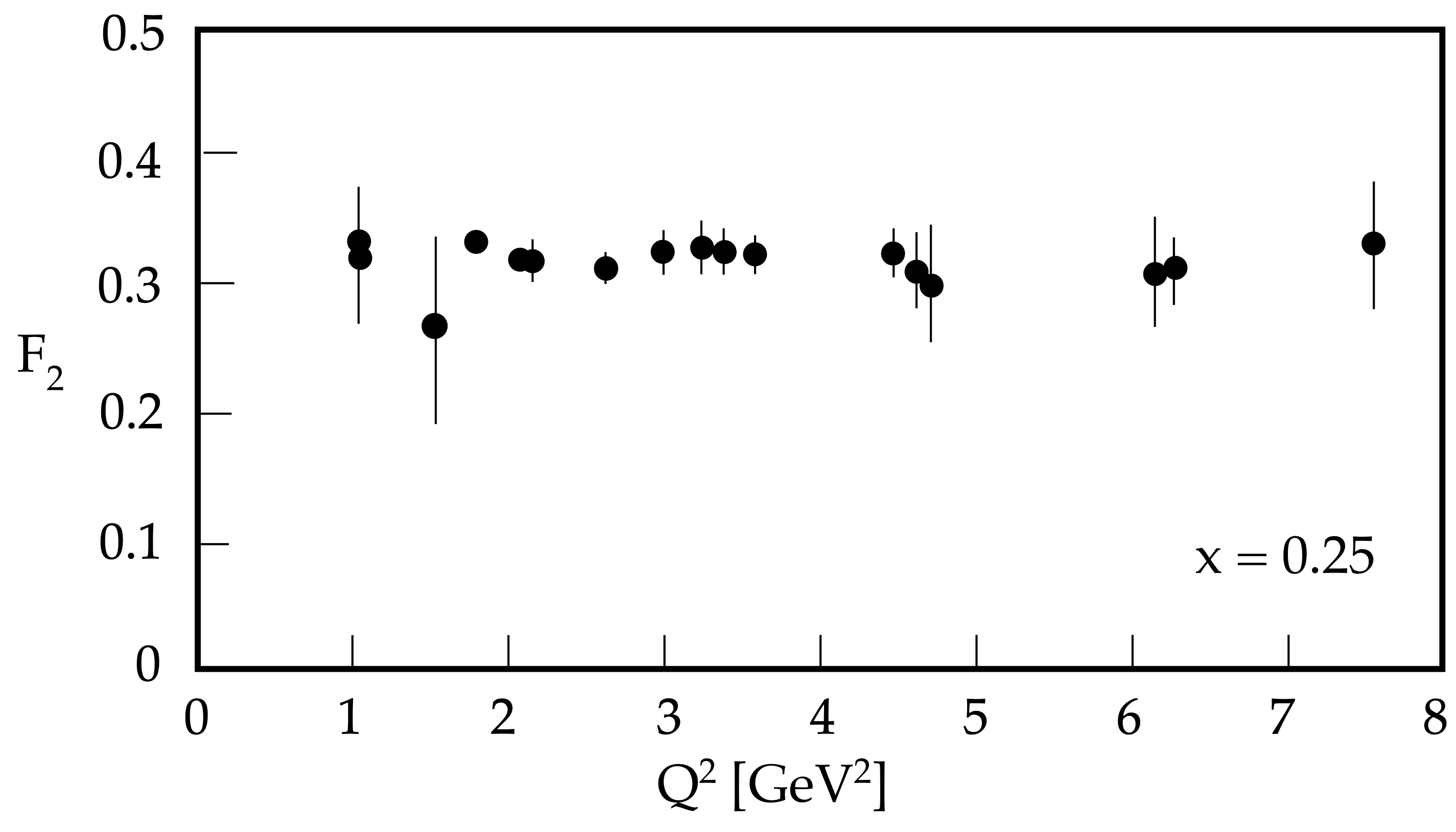}
\centering
\caption[]{$F_2$ as a function of $Q^2$  for $x$ = 0.25 (adapted from Ref.\cite{annurev:/content/journals/10.1146/annurev.ns.22.120172.001223}). 
\label{Scalingfig}}
\end{figure}

In the first experiment with the 20 GeV spectrometer  at $\theta =6^\circ$, the counting rates in the deep inelastic region were much higher than expected from eq. (\ref{Rosen}) \cite{Breidenbach:1969kd}. This was the first sign of scattering on hard constituents in the proton, analogous to the observation of the large angle $\alpha$ scattering on gold nuclei, which led to the discovery of the atomic nucleus  in 1911. Fig. \ref{Scalingfig} shows the behavior of $F_2$ at various scattering angles and incident energies as a function of $Q^2$ at a fixed $x$ = 0.25. The structure function does not depend on $Q^2$, which means that the size of the constituents cannot be determined by increasing $Q^2$ (and correspondingly decreasing the photon wavelength), because they are point-like.

\begin{figure}[htb]
\includegraphics[width=0.25\textwidth]{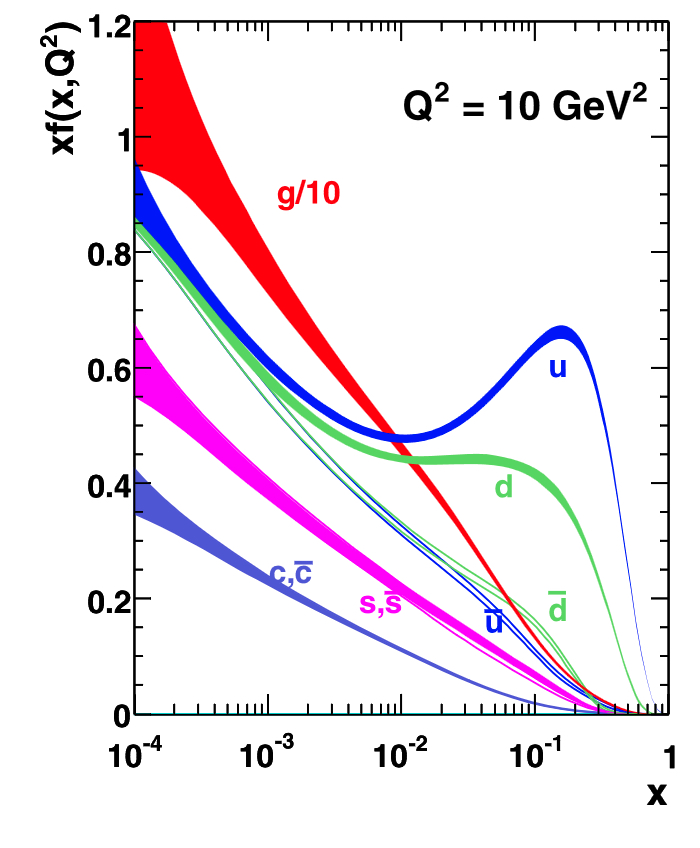}
\centering
\caption[]{Structure functions of the valence $u$ and $d$ quarks, sea quarks and antiquarks  ($s$, $\bar{s}$, $c$,  $\bar{c}$) and gluons $g$ as a function of $x$ for $Q^2$ = 10 GeV$^2$ \cite{Martin:2009iq}.
\label{Scaling4}}
\end{figure}

\noindent
$F_2(x)$  was found  \cite{annurev:/content/journals/10.1146/annurev.ns.22.120172.001223} to reach a constant value for $x\to 0$, which implied that some of the functions $q_i(x)$ had to grow to infinity, see eq. (\ref{eq:F2x}). Apart from the three `valence' quarks $uud$, there are gluons producing transient pairs of virtual quarks-antiquark pairs, the `sea' quarks and antiquarks, which dominate for small  $x$ values (Fig. \ref{Scaling4}). 

The structure function $F_2$ of the proton has been extensively measured  by many experiments with electrons, positrons and muons, in particular at the HERA electron-proton collider in Hamburg \cite{ParticleDataGroup:2014cgo}. Fig. \ref{Scaling4} shows the structure functions of the proton at $Q^2$ = 10 GeV$^2$. Scaling violation, {\it i.e.} a  strong dependence on $Q^2$, appears for low and high values of $x$ due to higher order processes (such as gluon emission by the struck quark).  The discovery of scaling was therefore a stroke of luck, reminiscent of the discovery of the atomic nucleus  with low energy  $\alpha$ particles that were scattered  without penetrating the nuclear core. 

\subsection{Deep inelastic neutrino-nucleus scattering}
\label{sec:NUQ}
Consider high energy muon neutrinos $\nm$ or antineutrinos $\anu$, scattering off a nucleon with mass $M$. Let $ E $ be the laboratory energy of the incoming $\nm$ (or $\anu$), $ E '$ that of the  outgoing muon, and $m = xM$ the mass of the struck quark. We denote by $y=\frac{E - E'}{E}$ the fraction of the incident energy carried out by the scattered quark (hence the emitted hadrons).

The neutrino  scatters only off a $d$ quark ($ \nm d\to \mu^-u$) and the antineutrino off a $u$ quark ($ \anu u\to \mu^+ d$).  For $\nm$ the angular distribution of the $\mu^-$ is isotropic in the center of mass system and, correspondingly, the cross section is independent of $E'$. With $\anu$ the outgoing $ \mu^+$ cannot be emitted backwards in the center of mass system ($E'=0$ in the lab) due to total helicity conservation. Denoting by  $\sigma$ and $\overline\sigma$ the cross sections  for $\nm$, respectively $\anu$, one gets (see {\it e.g.} Refs.\cite{HalzenMartin,10.1088/978-0-7503-1140-3})
\be
\frac{d\sigma}{dy} =   \frac{2 G^2 m E}{\pi} ,  \ \ \frac{d\overline\sigma}{dy} = \frac{2 G^2 m E}{\pi} (1 - y)^2.
\ee
where $G$ = $G_F\cos\theta_C$, with $G_F$ the Fermi coupling constant and $\theta_C\simeq 13^\circ$ the Cabibbo angle.
Ignoring sea quarks, the mass distribution  $q(x)$ of the struck quark is normalized to $\int_0^1 q(x) dx = 3$. The   differential cross sections for scattering off nuclei that  contain equal numbers of protons and neutrons (such as deuterium or $^{12}$C) are
\be
\frac{d^2\sigma}{dxdy} = \frac{1}{2}\cdot\frac{2G ^2 xME}{\pi} q(x) , \ \  \ 
\frac{d^2\overline\sigma}{dxdy} = \frac{1}{2}\cdot\frac{2G ^2xME}{\pi} q(x) (1 - y)^2. 
\ee
The factors  $\frac{1}{2}$ arise from the fact that the $\nm$ and $\anu$  scatter only off $d$ and $u$ quarks, respectively.  Integrating over $x$ leads to
\be
\frac{d\sigma}{dy}  = \frac{G ^2\overline{x}}{\pi} ME, \ \  \ 
\frac{d\overline\sigma}{dy} = \frac{G ^2\overline{x}}{\pi} ME (1 - y)^2
\label{eq:dsdy}
\ee
with $\bar{x} = \int_0^1 xq(x)dx$.
If the quarks carry the whole nucleon mass, then one quark carries on average $\frac{1}{3}$ of the nucleon mass, 
hence the average would be $\overline{x}$ = 1.
The total cross sections for deep inelastic scattering are obtained by integrating over $y$:
\be
\sigma  = \frac{G^2\overline{x}}{\pi} ME ,  \ \  \  \overline\sigma  = \frac{G ^2\overline{x}}{\pi} \frac{ME}{3} ,
\label{eq:sigtot}
\ee
leading to the ratio $\sigma/\overline\sigma = 3$ .

\begin{figure}[htb]
\includegraphics[width=0.30\textwidth]{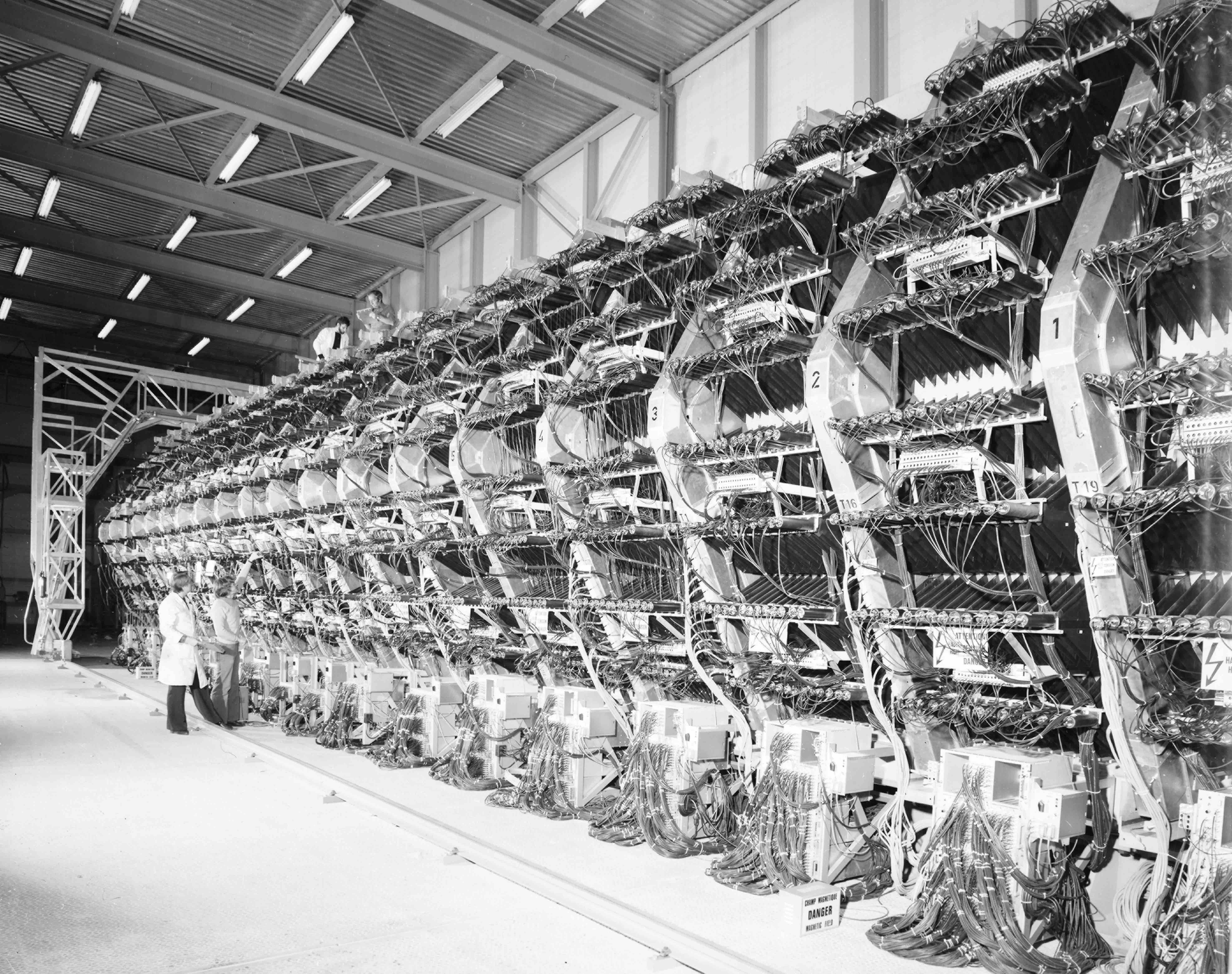}
\centering
\caption[]{The CDHS detector to study deep inelastic neutrino-nucleus scattering  \cite{deGroot:1978feq} (Photo credit CERN).
\label{CrosSecNeub}}
\end{figure}

The incident energy $ E $ and   $ y $ can be determined by  measuring the total energy deposited by the hadrons  and the muon energy $ E '$. Fig.\ref{CrosSecNeub} shows the neutrino detector of the CDHS collaboration at CERN \cite{deGroot:1978feq}. Hadrons of 200 GeV, produced by 400 GeV protons from the CERN SPS impinging on a target, were selected, focused and allowed to decay in a 300 m evacuated tube. Neutrinos with energies up to $\sim$200 GeV were produced from $\pi^+(K^+)\to\mu^+\nm$ decay by focusing  the positively charged mesons, antineutrinos from  $\pi^-(K^-)\to\mu^-\anu$ by focusing the negatively charged ones. A 172 m long steel shield absorbed the surviving mesons. The detector  consisted of 19 modules comprising thick steel plates as neutrino targets (Fe contains nearly as many protons as neutrons). The plates were magnetized to determine with drift chambers  the momentum of the outgoing muon. The energy of the induced hadron shower was measured with plastic scintillators.

\begin{figure}[htb]
\includegraphics[width=0.48\textwidth]{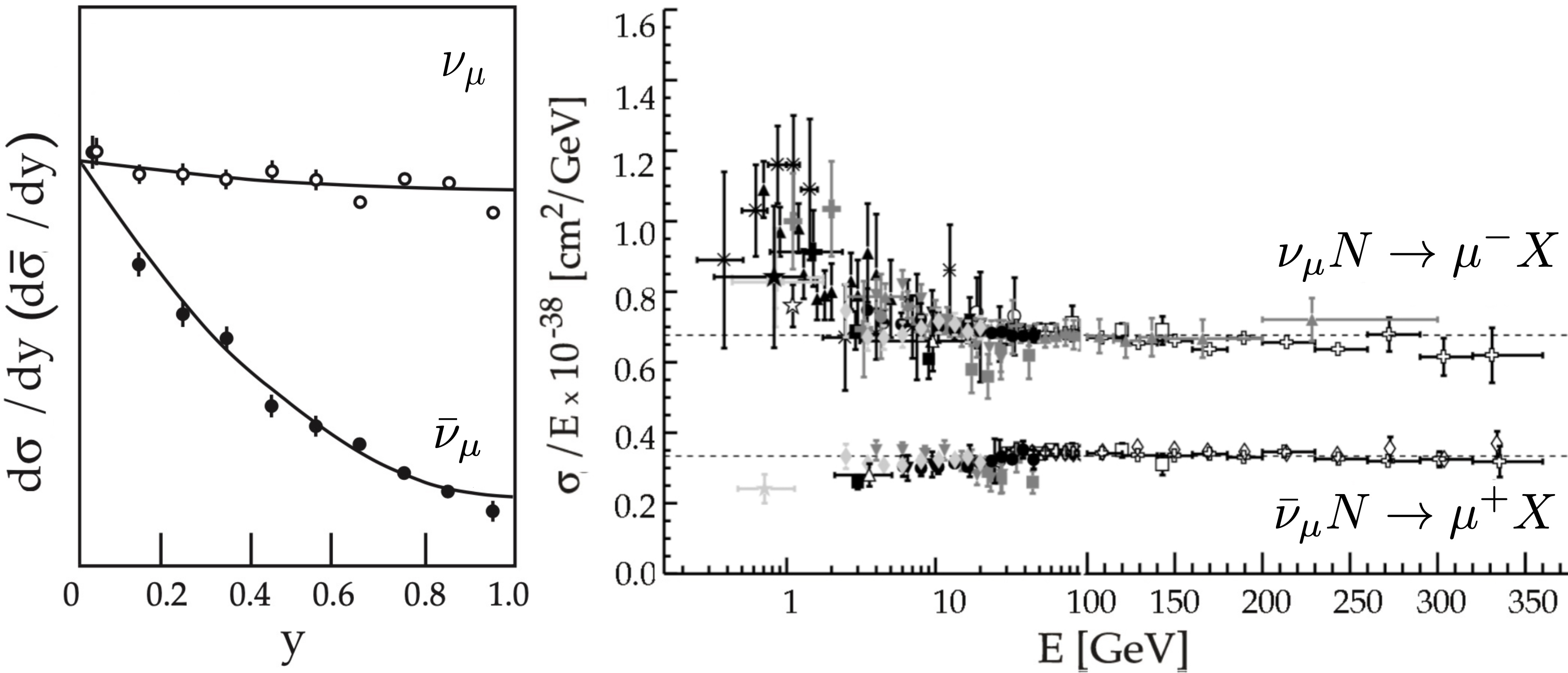}
\centering
\caption[]{Left: $\nm N \to \mu^- X$ and $\anu N \to \mu^+ X$  cross sections as a function of $y$ (arbitrary units) with fits of the form $a + b (1 - y)^2$ \cite{deGroot:1978feq}. Right: Total cross sections divided by $E$ from various experiments  versus laboratory energy $E$ \cite{ParticleDataGroup:2014cgo}. 
\label{CrosSecNeua}}
\end{figure}

Fig. \ref{CrosSecNeua} (left)   shows the differential cross sections $d\sigma/dy$ and  $d\overline\sigma/dy$.  For neutrinos the cross section is almost constant, decrea\-sing slightly as a function of $y$, due to sea $q\bar{q}$ pairs.  Neutrino-antiquark scattering is equivalent to antineutrino-quark scattering  (by $ CP$ invariance) and contributes a small admixture of the term $ (1 - y)^2 $ to the cross section. The measurements agree with  eq. (\ref{eq:dsdy}), consistent with the presence of spin $\frac{1}{2}$ scattering centers in the nucleon.

At high energies the total cross sections  grows  proportionally to the laboratory incident energy $E$ (Fig. \ref{CrosSecNeua}, right) as expected, but the ratio $\sigma/\overline\sigma \simeq 2$   differs from the expected value of  3, due to $q\bar{q}$ sea pairs induced by gluons.   The contribution from sea antiquarks to $\sigma$ is $\frac{1}{3}\epsilon$, that from sea quarks to $\overline\sigma$ is 3$\epsilon$, where $\epsilon$ is the fraction of sea antiquarks in the nucleon, therefore $\sigma/\overline\sigma = 3 (1 +\epsilon/3)/(1 + 3 \epsilon) \simeq 2$ and hence $\epsilon \simeq$ 20\%. The contribution $\overline{x}$ of quarks to the nucleon mass can be also estimated from the  cross-sections in Fig. \ref{CrosSecNeua} (right). One finds with (\ref{eq:sigtot}) $\overline{x} \simeq$  0.43, which shows that roughly half of the nucleon mass is carried by quarks, the other half by gluons.

\begin{figure}[htb]
\includegraphics[width=0.20\textwidth]{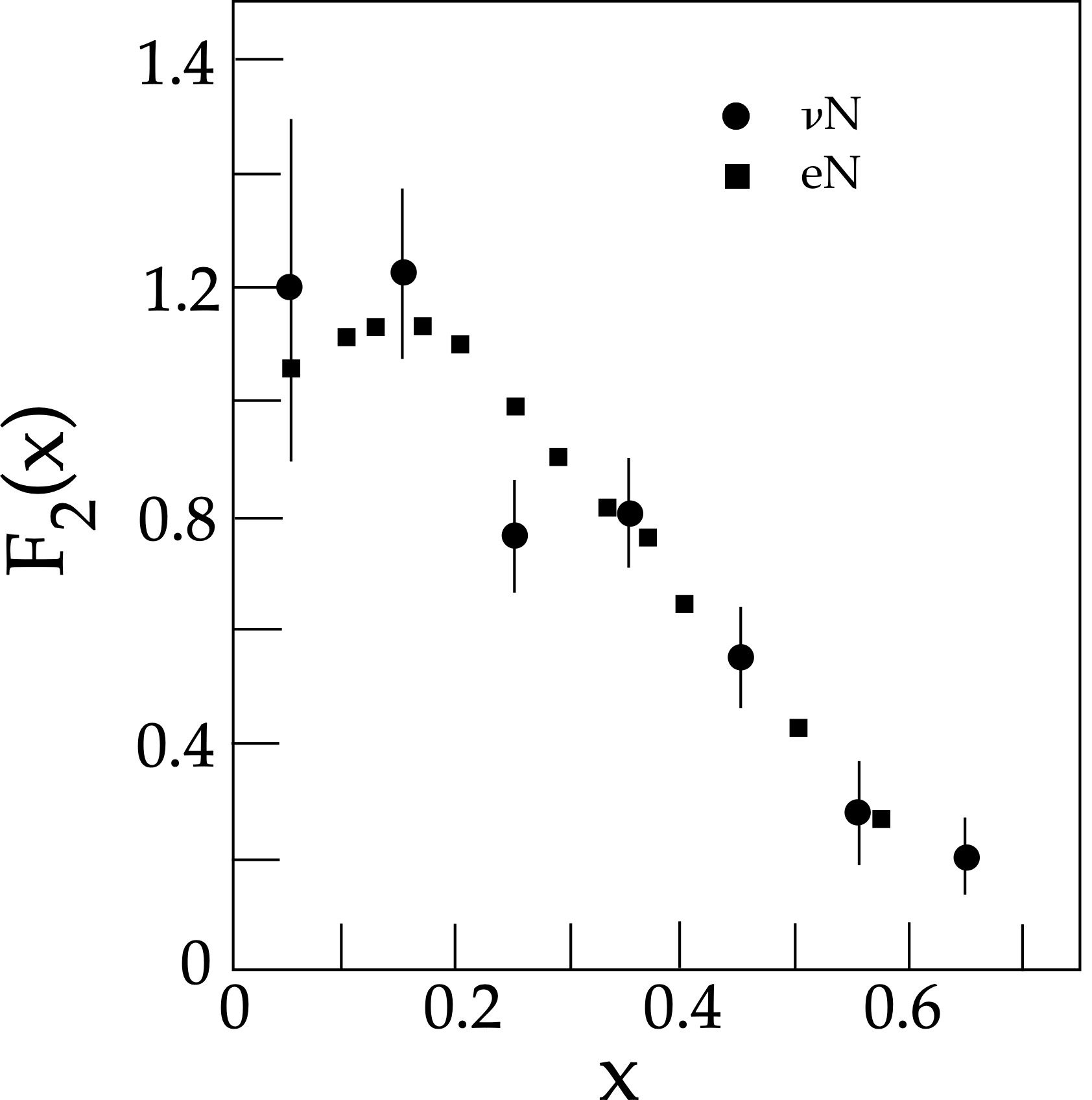}
\centering
\caption[]{Comparison of  deep inelastic neutrino-nucleon with deep inelastic electron-nucleon scattering (adapted from Ref.\cite{doi:10.1126/science.256.5061.1287}).
\label{Neutelec}}
\end{figure}

Finally, let us compare the  structure functions for deep inelastic neutrino with deep inelastic electron scattering. The structure function $F_2(x)$ from neutrino data taken with the Gargamelle heavy-liquid BBC at CERN \cite{Budagov:1969at}, and from $e^-p$ and 
$e^-n$ scattering at SLAC are plotted in Fig. \ref{Neutelec}. The electron data have been divided by 5/18 (the average squared quark charge between proton and neutron) to compare with neutrino data. The agreement between electron and neutrino scattering is excellent, thus showing that both probes see the same point-like spin $\frac{1}{2}$ objects in the nucleon. Furthermore, the fractional quark charges are confirmed. 

\subsection{Quark and gluon jets}
\begin{figure}[htb]
\includegraphics[width=0.48\textwidth]{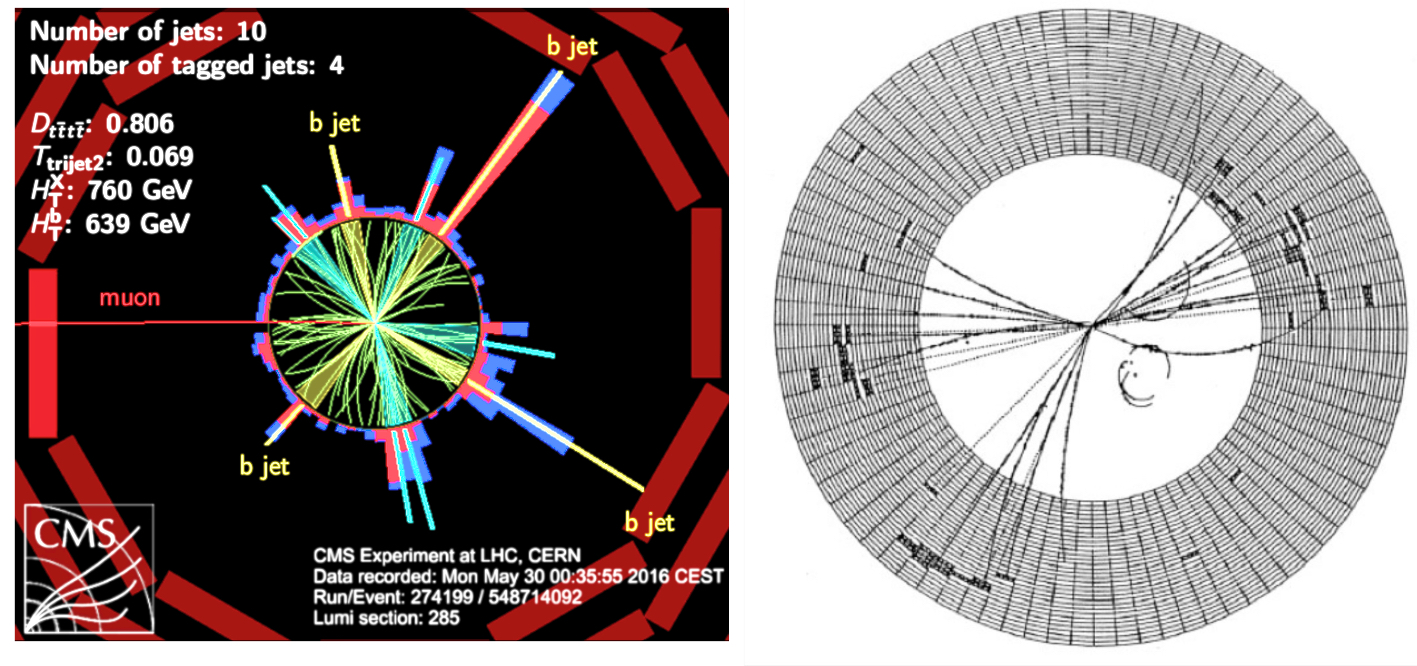}
\centering
\caption[]{Left: a multi-jet event including 4 b-quark jets observed at 13 TeV by CMS at the CERN LHC. The colored histogram shows the energy deposits in the calorimeters (Image Credit: CERN/CMS Collaboration.)
Right: a three-jet event registered by JADE in  31 GeV $e^+e^-$ collisions at PETRA \cite{Soding:2010zz}.
\label{Jets}}
\end{figure}

The discovery of hadron jets in high energy interactions  support the parton model, providing direct evidence for the existence of constituents in hadrons. The hadron jets are produced by  the fragmentation of the original parton and fly roughly in the same direction. Fig.  \ref{Jets} (left) shows jets produced in $pp$ collisions at the LHC and Fig. \ref{Jets} (right) a 3-jet event generated in $e^+e-$ collisions at the PETRA collider at DESY. One of the original quarks has emitted  a gluon which turned into a $\qqbar$ pair to produce a third hadron jet. The spin $ J = 1$ of the gluon was determined by TASSO at PETRA from the angular correlations between the jets \cite{TASSO:1980lqw}. 

\section{Experimental evidence for color}
\label{sec:colour}
Before the introduction of color  the spin $\frac{3}{2}$  baryons ($\Delta^{++}$ = $uuu$, $\Delta^{-}$ = $ddd$ and $\Omega^-$ = $sss$) appeared to violate the Pauli principle, which required the wave functions of identical fermions to be antisymmetric under permutations. The introduction of three `colors' (say red, green and blue) for each flavor, solved the conundrum. But color would increase the number of observed hadrons, therefore only `colorless' hadrons should exist. Hence free quarks are also not observed:
 Free quark searches were performed without success by looking for fractional charges with various methods, such as ionization (dE/dx) measurements at colliders, accelerators,  and in cosmic rays, or by studying nuclear fragments in heavy ion collisions with etched CR-39 plastic foils. For a review see Ref.\cite{Perl:2004qc}. 

\begin{figure}[htb]
\includegraphics[width=0.25\textwidth]{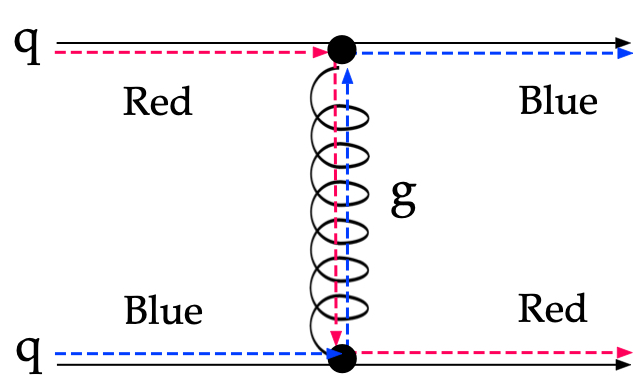}
\centering
\caption[]{Quarks interact through the exchange of colored gluons.
\label{Colour}}
\end{figure}

A  colorless baryon consists  of a superposition of the six permutations of color. Antiquarks  come in three anticolors and a meson is a linear superposition of quarks with colors and antiquarks with the corresponding anticolors. 
The interaction between quarks is mediated by gluons through the exchange of color (Fig. \ref{Colour}), which is conserved, and the underlying symmetry  is SU(3)$_{\rm c}$.
Hence a gluon consists of  a color and an anticolor. Nine gluons can be  constructed from three colors and three anticolors but one is colorless. Since free gluons are not observed, only eight gluons exist which are the gauge fields of QCD. The concept of three quark `colors', leading to the development of QCD in the 1970s,  dates back to  papers by Greenberg \cite{Greenberg:1964pe} and Han \& Nambu \cite{Han:1965pf}. 

\begin{figure}[htb]
\includegraphics[width=0.30\textwidth]{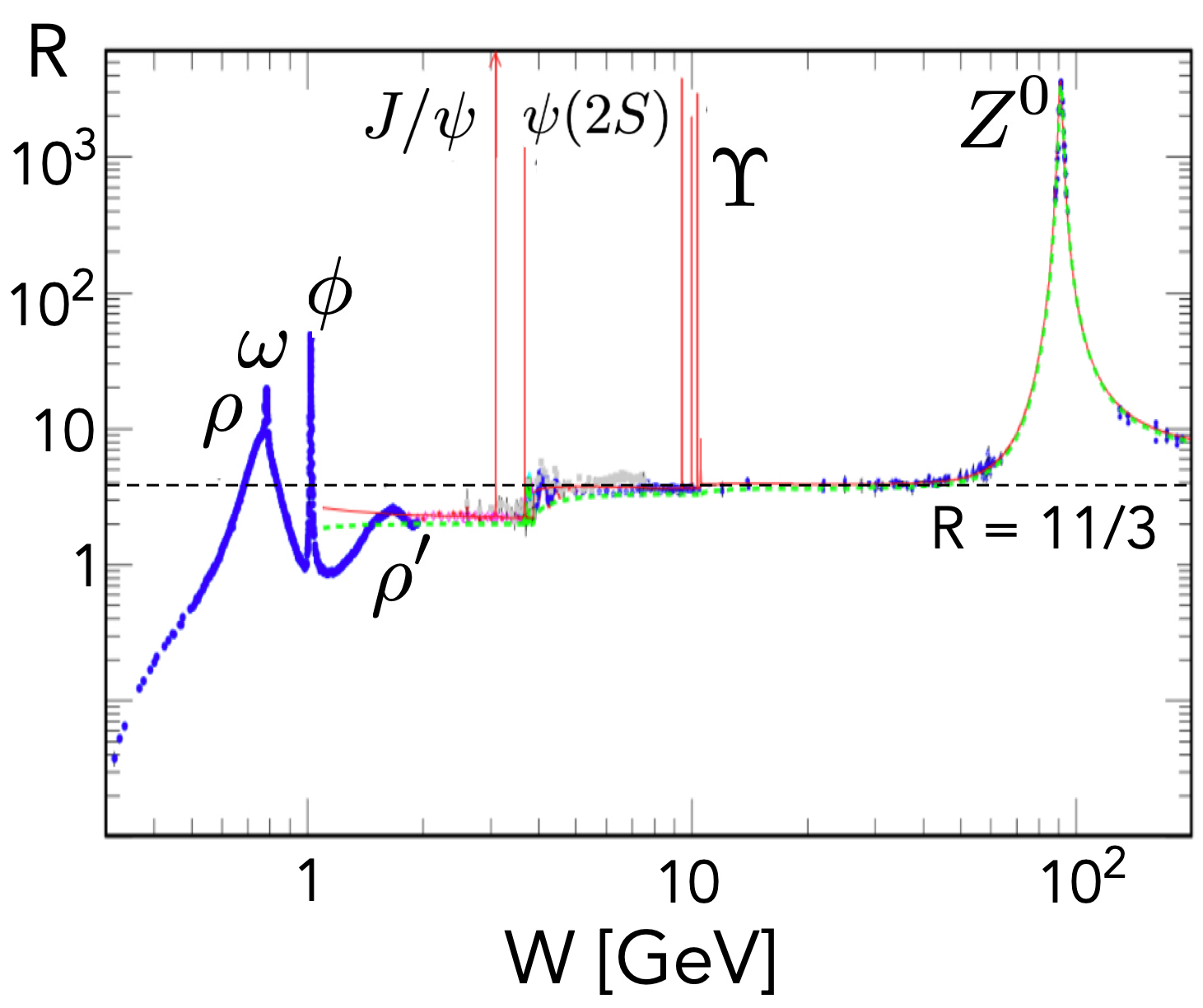}
\centering
\caption[]{$e^+e^-$ annihilation cross section into hadrons, divided by the calculated non-resonant cross section $\sigma(e^+e^-\to\gamma\to\mu^+\mu^-)$  \cite{ParticleDataGroup:2014cgo}. 
\label{Colored}}
\end{figure}

Color has been established by several experimental facts, in particular in the production of quark-antiquark pairs $\uubar$, $\ddbar$, $\ssbar$ and $\ccbar$, $\bbbar$ (sections \ref{sec:charm} and \ref{sec:bottomq}) studied in high energy $e^+e^-$ collisions, which  proceed  through the virtual intermediate photon  $e^+e^-\to\gamma\to q\bar{q}$, followed by  fragmentation into hadrons. Fig. \ref{Colored} shows the ratio $R$ of the production cross section, divided by the calculated non-resonant electromagnetic cross section for $e^+e^-\to\mu^+\mu^-$, as a function of center-of-mass energy $W$. The various vector mesons  series, as well as the $Z^0$ boson, are observed above non-resonant hadron production. 
The ratio $R$ of non-resonant cross sections is given by
\be
R = \frac{\sigma(e^+e^-\to \ {\rm hadrons})}{\sigma(e^+e^-\to 
\mu^+\mu^-)} = \frac{\alpha^2\sum q_i^2}{\alpha^2},
\label{Rvalue}
\ee
where $\alpha$ is the fine structure constant (we have ignored a small  QCD correction). The numerator sums over all quark charges $q_i$ and 
the phase space factors cancel between numerator and denominator since at sufficiently high energies all masses can be neglected. Above   the $b\overline {b}$ threshold of around 10 GeV one would expect 
$R =   \left(\frac{2}{3}\right)^2+\left(-\frac{1}{3}\right)^2+
\left(-\frac{1}{3}\right)^2+\left(\frac{2}{3}\right)^2+\left(-\frac{1}{3}\right)^2= \frac{11}{9}$, in contradiction with the measured $R = 11/3$  (Fig. \ref{Colored}). The missing factor of $3$ is the smoking gun evidence for the existence of three quark colors. 

The measured probability for the decay of a $W^+$ boson into $e^+\nl$, $\mu^+\nm$ and $\tau^+\nt$ is 11\% each (the W boson being very heavy, quark masses can be neglected). One expects  $W^+\to u\overline{d}$ and $c \overline{s}$ to contribute each also 11\% ($W^+\to u \bbar$, $u\sbar$, $ c \dbar$ are  `Cabibbo' suppressed). Three quark colors are needed to reach 100 \%. 

\begin{table}[htb]
\begin{center}
\caption[]{The storage/collider rings mentioned in this review. The last column gives the maximum energy per beam in GeV. \\$^\dagger$ Nominal energy.}
\footnotesize
\begin{tabular}{l l l l l}
\hline
ADONE & Frascati & $e^+e^-$ & 1969--1993 & 1.5 \\
SPEAR (-II) & SLAC & $e^+e^-$ & 1972--1990 & 4.2 \\
DORIS & DESY &$e^+e^-$ & 1974--1992 & 5.3 \\
VEPP-2 (2M) & Novosibirsk & $e^+e^-$ & 1967--2000 & 0.7 \\
PETRA & DESY & $e^+e^-$ & 1978--1986 &19 \\
CESR (-C) & Cornell & $e^+e^-$ & 1979--2008& 6 \\
$S\!p\bar{p}S$ & CERN & $\bar{p}p$ & 1981--1991& 315 \\
LEAR & CERN &$\bar{p}$ & 1982--1996 & 0.005--1.3 \\
TEVATRON & Fermilab &$\bar{p}p$ & 1987--2011 & 980 \\
BEPC (-II) & Beijing & $e^+e^-$ & 1989-- & 4 \\
LEP& CERN &  $e^+e^-$ & 1989--2000 & 104.6 \\
VEPP-4M & Novosibirsk & $e^+e^-$ & 1994-- & 6\\
KEKB & KEK & $e^+e^-$ & 1999--2010 & 8$^\dagger$($e^-$), 3.5$^\dagger$($e^+$)\\
DA$\Phi$NE& Frascati & $e^+e^-$ & 1999-- & 0.51 \\
PEP-II & SLAC &  $e^+e^-$ & 1999--2008 & 9$^\dagger$($e^-$), 3.1$^\dagger$($e^+$)\\
LHC& CERN &  $pp$ & 2009-- & 7000 \\
\hline
\end{tabular}
\label{tab:Facilities}
\end{center}
\end{table}

\section{The November revolution}
\label{sec:charm}
\begin{figure}[htb]
\includegraphics[width=0.48\textwidth]{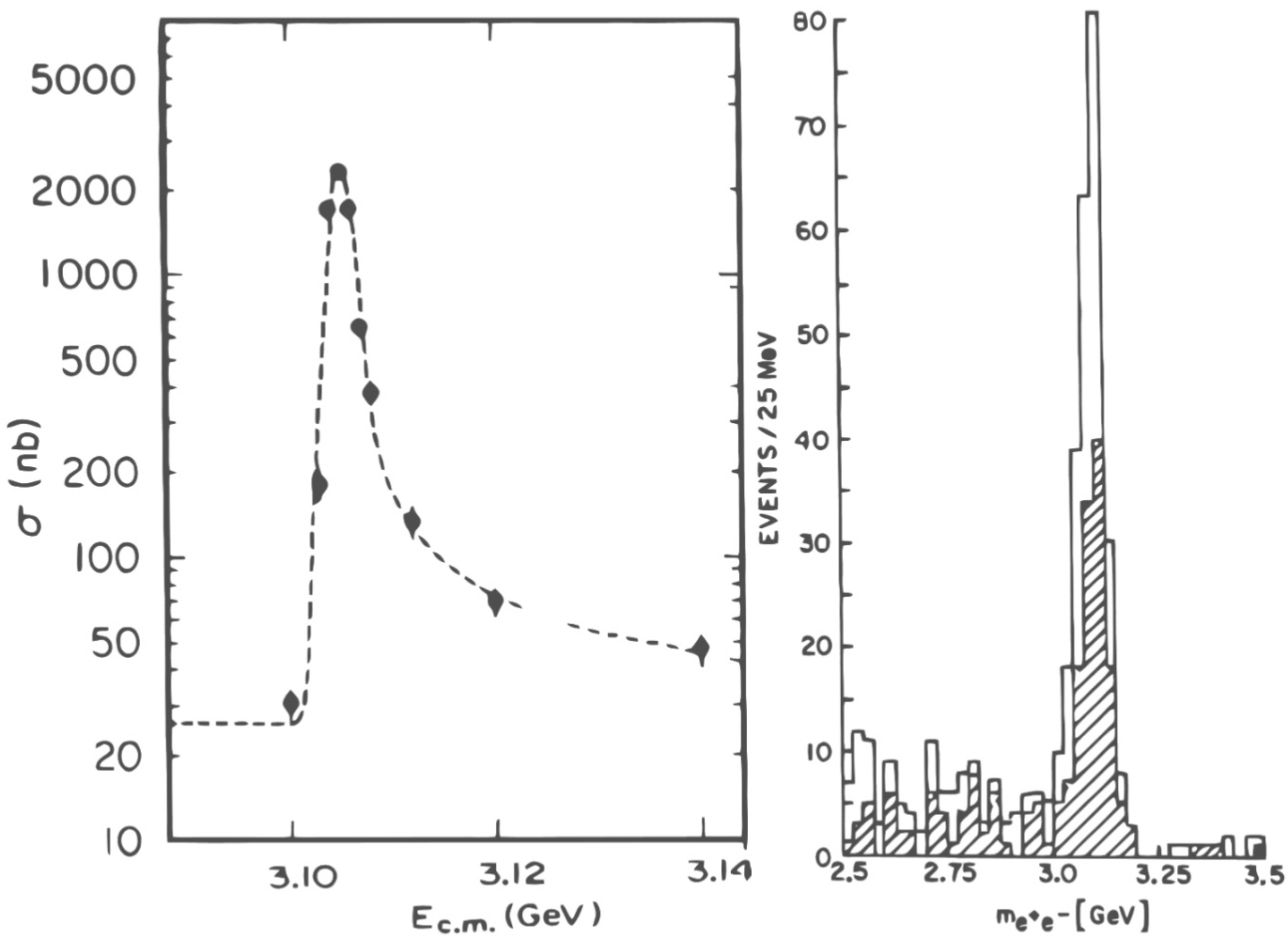}
\centering
\caption[]{Left: The $\psi$ resonance decaying into hadrons  in $e^+e^-$ collisions as a function of center of mass energy  \cite{SLAC-SP-017:1974ind} (the long tail is due to radiative processes). Right: $e^+e^-$ invariant mass distribution in $p$Be interactions showing the $J$  resonance for two spectrometer current settings \cite{E598:1974sol}.  
\label{JpsiDiscovery}}
\end{figure}

In the early seventies the ratio $R$  in  Fig. \ref{Colored} was found  to be larger above $\sim$ 3 GeV than the expected $R$ = 2 with three quark flavors. This was one of the motivations to build the SPEAR storage ring at SLAC (Table \ref{tab:Facilities}).  The cylindrical magnetic spectrometer MARK I, installed around the $e^+e^-$ collision point, was equipped with wire spark chambers, scintillation and shower counters. 
In November 1974, while scanning the collision energy in small steps, a sharp resonance was observed  around 3.1 GeV to decay into
hadrons and $\mu^+\mu^-$ pairs  (Fig. \ref{JpsiDiscovery}, left).  
The resonance was also discovered with 29 GeV protons from the AGS at BNL, impinging on a beryllium target and decaying into $e^+e^-$ pairs (Fig. \ref{JpsiDiscovery}, right). The two leptons were separated from other hadrons by \v{C}erenkov hydrogen gas counters and momentum analyzed by two magnetic spectrometers. The new reso\-nance was named $\psi$ at SLAC and $J$ at BNL, and was immediately confirmed at the ADONE storage ring in Frascati \cite{Bacci:1974za}. 

\begin{figure}[htb]
\includegraphics[width=0.40\textwidth]{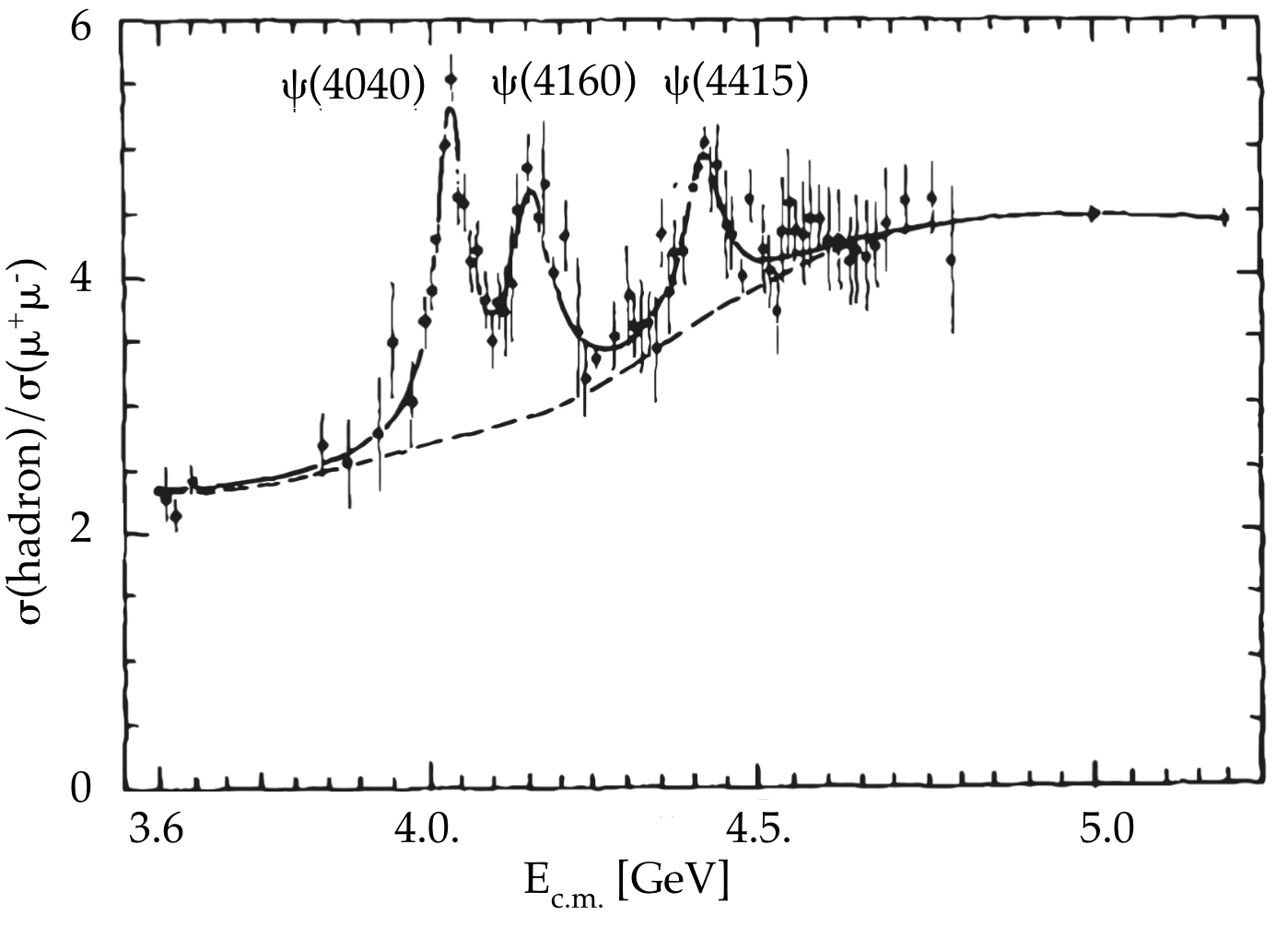}
\centering
\caption[]{Cross section for hadron production normalized to $\gamma\to\mu^+\mu^-$ in $e^+e^-$ collisions, as a function of center of mass energy, showing three $\psi$ excitations \cite{DASP:1978dns}.  
\label{Threepsi}}
\end{figure}

The quantum numbers $J^{PC} = 1^{--}$ follow from the observed interference between  $e^+e^-\to\psi\to\mu^+\mu^-$ and $e^+e^-\to\gamma\to\mu^+\mu^-$ channels \cite{Boyarski:1975ci}. The measured width  was determined by the energy spread of the colliding beams at  the MARK I detector ($<$1.3 MeV) and by spectrometer resolution at the AGS.  The natural width $\Gamma$ can be estimated by  integrating the cross section in Fig. \ref{JpsiDiscovery} (left) and assuming a Breit-Wigner resonance shape:

\be
\int_{-\infty}^{\infty} \sigma dE =  \frac{3}{4}\cdot\frac{8\pi^2}{M^2} f(e^+e^-) f(h)\Gamma ,
\ee
for a spin 1 resonance of mass $M$, and where  $f(e^+e^-)$and $f(h)$ are the decay branching fractions for $\psi\to e^+e^-$ ($\sim$6\%) and  $\psi\to$ hadrons ($\sim$88\%), respectively. The width of the $\psi$ ($\Gamma$ =  93 keV) is surprisingly small.  Another narrow state, the $\psi'$ ($\Gamma$ = 303 keV),  was detected by Mark I in  $e^+e^-$ collisions, about 600 MeV above the $J/\psi$ \cite{Abrams:1975zp}, followed a few years later by a much broader resonance, the $\psi(3770)$ with a width of 27 MeV \cite{Rapidis:1977cv}.   Further broad states  (Table \ref{tab:psiups}) were observed at higher center of mass energies with the DASP detector (Fig. \ref{Threepsi}) at  the $e^+e^-$ DORIS  ring. DASP was a double arm spectrometer, equipped with proportional wire chambers, analyzing magnets, spark chambers, time of flight and shower counters, as well as steel absorbers to detect muons. 

\begin{table}[htb]
\begin{center}
\caption[]{The left columns list the  $\ell$ = 0  radial $J^{PC} = 0^-(1^{--})$   $\ccbar$ states, the right columns the $\bbbar$ ones (section \ref{sec:quarkonium}). The $\psi'$  is now called $\psi(2S)$. The $\psi(3770)$ and $\psi(4160)$ are not shown, they are $\ell$ = 2 orbitals  ($1^3D_1$ and $2^3D_1$). Masses and widths are from Ref.\cite{ParticleDataGroup:2026cfk}.}
\footnotesize
\begin{tabular}{llllll }
\hline
 & $M$ (MeV) & $\Gamma$ (MeV)& & $M$ (MeV)& $\Gamma$ (MeV)\\
\hline
$J/\psi(1S)$ & 3097 & 0.093 & $\Upsilon(1S)$ &  \,\,\,9,460 & 0.054 \\
$\psi(2S)$ & 3686 & 0.294 & $\Upsilon(2S)$ & 10,023 & 0.032 \\
$\psi(3S)$ & 4040 & 84 & $\Upsilon(3S)$ & 10,355 & 0.020 \\
 $\psi(4S)$ & 4415  & 110  & $\Upsilon(4S)$ & 10,579 & 21 \\
&  & & $\Upsilon(5S)$ & 10,885 & 37 \\
\hline
\end{tabular}
\label{tab:psiups}
\end{center}
\end{table}

The reason for the  small $J/\psi$ and $\psi'$ widths  is  the existence of the charm quark $c$\footnote{The first appearance of the charm quark dates back to 1971  when a heavy particle decay was observed in a nuclear emulsion exposed to cosmic rays \cite{2008PJA8401B-03}.  An unexplained shoulder had also been observed at the AGS in the  $\mu^+\mu^-$ invariant mass in hadron collisions, which could not be resolved due to insufficient energy resolution \cite{Christenson:1970um}. 
}.  This  fourth quark with charge  $\frac{2}{3}$ had been  postulated in 1970 to explain the absence of flavor changing neutral currents, the $d\leftrightarrow s$  transitions (GIM mechanism \cite{Glashow:1970gm}). The $J/\psi$ and $\psi'$ are  bound states of pairs of charm-anticharm ($c\bar{c}$) quarks which cannot fragment into pairs of `charmed' mesons ($q\bar{c}$ and $\bar{q}c$, with $q=u,d,s$), because their  masses are smaller than the sum of the two charmed meson masses. They decay electromagnetically to lepton-antilepton pairs or to light hadrons via the emission of gluons which are suppressed\footnote{The Okubo-Zweig-Iizuka (OZI) rule states that strong interactions without quark transfer to a final state hadron are forbidden.}, leading to a long lifetimes and hence to narrow widths.  

\subsection{Charmed hadrons}
\begin{figure}[htb]
\includegraphics[width=0.48\textwidth]{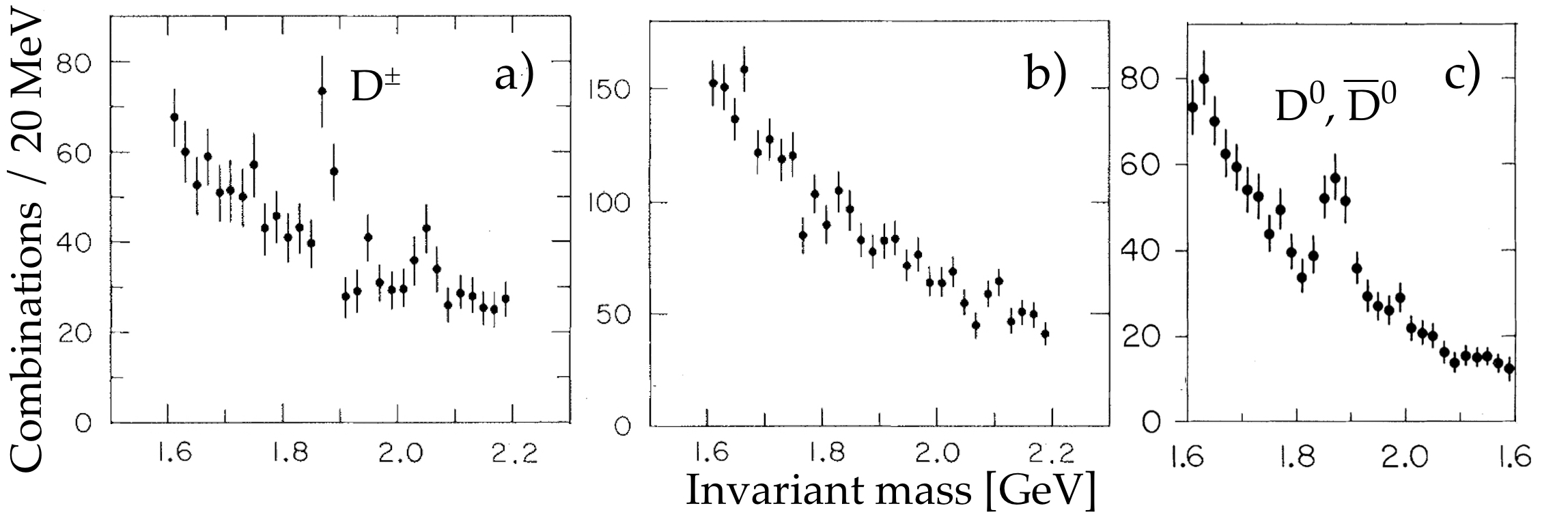}
\centering
\caption[]{a) $K^\mp\pi^\pm\pi^\pm$ mass distribution, showing evidence for the charged $D(1870)^\pm$ \cite{Peruzzi:1976sv}; b) The wrong charge distribution $K^\pm\pi^+\pi^-$ does not show any signal; c) $\pi^\pm K^\mp$ mass distribution showing the $D^0 $ (and $\bar{D}(1865)^0$) \cite{Goldhaber:1976xn}.
\label{Ddiscovery}}
\end{figure}

\begin{figure}[htb]
\includegraphics[width=0.40\textwidth]{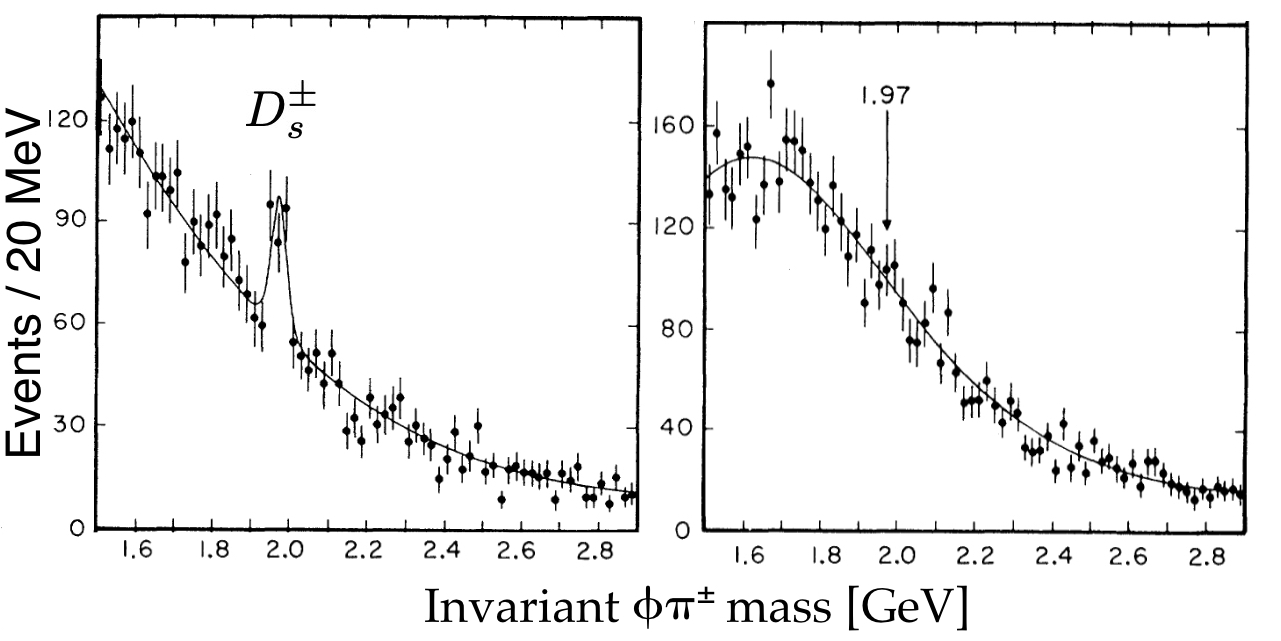}
\centering
\caption[]{Left: Observation of the $D_s(1968)^\pm\to\phi\pi^\pm$  (formerly called $F^\pm$). Right: distribution outside the $\phi\to K^+K^-$ signal \cite{CLEO:1983dsb}. 
\label{Dsdiscovery}}
\end{figure}

The threshold for the production of  a pair of charmed mesons lies between  the masses of the $\psi'$ and the $\psi(3770)$. Fig. \ref{Ddiscovery} shows the discovery of the $D$-mesons, the $D^\pm$ ($c\bar{d}$ and $\bar{c}d$) and  $D^0$  and $\bar{D}^0$ ($c\bar{u}$ and $\bar{c}$u) by MARK I. The  observation of the $D^\pm_s$ ($c\bar{s}$ and $\bar{c}s$)  by CLEO at the CESR storage ring in Cornell is shown in Fig. \ref{Dsdiscovery}. 

The $D$ mesons decay mostly into kaonic final states, due to  the transition $c\to sW^+$ (Cabibbo) favored over $c\to dW^+$, hence $D^\pm$ decays to $K^\mp$.  The lifetime of the $D$ meson is typically 1 ps \cite{ParticleDataGroup:2026cfk}, which is long enough to separate production and decay vertices in  experiments.

The $D$ and $D_s$ are $J^P = 0^-$ states. Akin to the $K^*$ there are also   $J^P = 1^-$ mesons. The $D^{*0}$, $\bar{D}^{*0}$, and $D^{*\pm}$  were observed at SPEAR \cite{Goldhaber:1977qn} by measuring the mass distribution recoiling against the $C$-conjugated $D$ at the $\psi(3S)$ and $\psi(4S)$. Likewise, the $D_s^{*\pm}$ was observed by DASP from the mass recoiling against the  $D_s^\mp$ \cite{DASP:1978gcx}. The mass differences between $D$ and $D^*$ (and between $D_s$ and $D_s^*$) are very close to the pion mass.  Since the $D^{*\pm}$ decays  to $\pi^+D^0(\pi^-\bar{D}^0)$, the charge of the very slow pion is useful to determine the charge of the decaying $D^{*\pm}$ and the ($c$ or $\cbar$) flavor of the associated neutral $D$.  The $D^{*\pm}$ also decays to $\pi^0D^\pm$ and $\gamma D^\pm$, the $D^{*0}$ to $\pi^0 D^0$ and $\gamma D^0$, but is not heavy enough for  $\pi^-D^+$. 

Several radial and orbital excitations of the $D$ and $D_s$ mesons have been observed \cite{ParticleDataGroup:2026cfk}. 
Of particular interest are the $D_{s0}^*(2317)^\pm$  ($J^P=0^+$) and $D_{s1}(2460)^\pm$  ($J^P=1^+$) which have been  produced in $e^+e^-\to \Upsilon(4S)$ (section \ref{sec:bottomq}) by BABAR at PEP-II \cite{BaBar:2006eep} and Belle at KEKB \cite{Belle:2003kup} (a drawing of the Belle detector is shown in Fig. \ref{BelleDet} below). They are much lighter than expected and decay isospin violating to $D_s^\pm\pi^0$, respectively $D_s^{*\pm}\pi^0$, hence are narrow ($<$4 MeV). They are candidates for four-quark states or $DK$ ($DK^*$) molecules.

\begin{figure}[htb]
\includegraphics[width=0.35\textwidth]{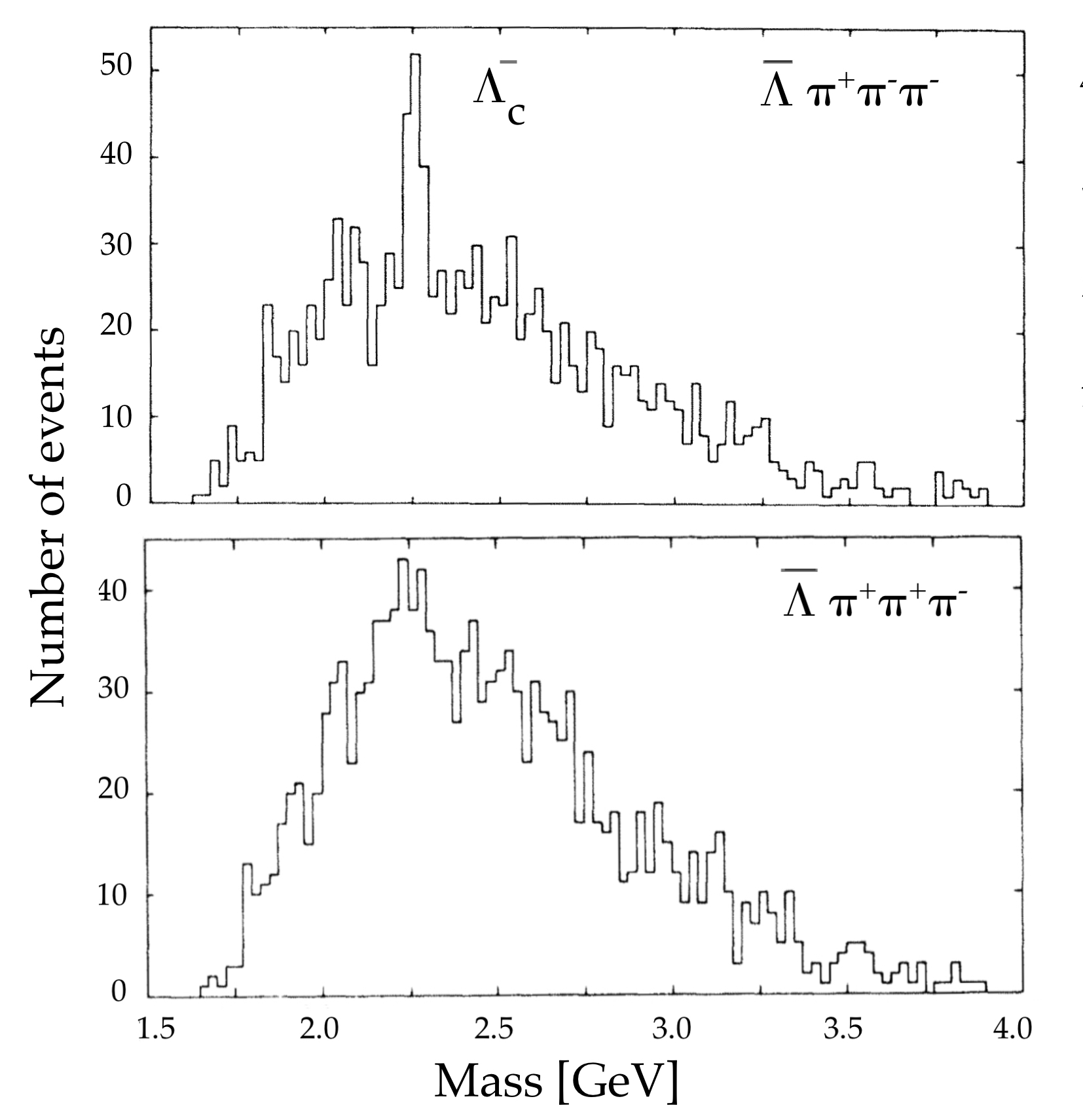}
\centering
\caption[]{Top: $\bar{\Lambda}\pi^+\pi^-\pi^-$  mass distribution revealing the charmed antibaryon $\Lambda_c^-$ (top). Bottom: $\bar{\Lambda}\pi^+\pi^+\pi^-$ mass distribution  \cite{Knapp:1976qw}.
\label{Lambdacfig}}
\end{figure}

The first charmed (anti)baryon was observed with a photon beam at Fermilab \cite{Knapp:1976qw}. Fig. \ref{Lambdacfig} (top) shows the $\bar{\Lambda}\pi^+\pi^-\pi^-$  mass distribution. The peak is due to an antibaryon with negative charge, which cannot be a strange antibaryon ($\bar{s}\bar{u}\bar{u}$) since no corresponding positively charged state  ($\bar{s}\bar{d}\bar{d}$) is observed in $\bar{\Lambda}\pi^+\pi^+\pi^-$ (bottom). The peak is due to the anti-$\Lambda_c^+$ ($\bar{c}\bar{u}\bar{d}$, called $\Lambda_c^-$). As rea\-lised later, a charmed baryon had been produced earlier in $\nm p \rightarrow \mu^-\Lambda \pi^-\pi^+\pi^+\pi^+$, seen in a  BBC exposure  at BNL \cite{Cazzoli:1975et}.  The reaction seemed to violate the $\Delta S = \Delta Q$ rule, but was instead due to the charmed baryon $\Sigma_c(2454)^{++}$ ($cuu$) produced  in $\nm p \rightarrow \mu^-\Sigma_c^{++}(\to \Lambda_c^+\pi^+ \to \Lambda\pi^-\pi^+\pi^+\pi^+$). The most precise measurement of its mass was obtained by producing the $\Sigma_c^{++}$ at the $\Upsilon(4S)$ and reconstructing its decay to $\Lambda_c^{+}(\to pK^-\pi^+)\,\pi^+$  with the Belle detector \cite{Belle:2014fde}. 

Many charmed baryons and their excitations have been observed \cite{ParticleDataGroup:2026cfk}, even a doubly charmed one by LHCb, the $\Xi_{cc}(3622)^{++}$  \cite{LHCb:2017iph}.

\subsection{Resonance depolarization}
\label{sec:resdepol}
To determine  the  masses of narrow resonances such as the $J/\psi$, one needs to calibrate the  energies of the $e^\pm$ beams very accurately. This is achieved by resonance depolarization:  In the ring the electrons become polarized through synchrotron radiation, with spins antiparallel to the magnetic field (the polarisation reaching at most 92\%). A radial magnetic field of frequency $\omega_d$, applied at a given location around the ring,  induces a radial component of the spin which then precesses around the vertical axis with the frequency $\Omega$ given by
\begin{equation}
\Omega = \omega_0\left( 1+ a\frac{E}{m_e}\right),
\label{eq:Omega}
\end{equation}
where $\omega_0$ is the cyclotron frequency and $a = (g-2)/2$ the anomalous magnetic moment of the electron.
When $\omega_d$ is in phase with the precession frequency around the ring, the spin  deviates more and more from the vertical direction with the number of revolutions, and the polarization is destroyed. This occurs when $\Omega$ fulfills the condition  $\Omega = n\omega_0 + \omega_d$ with $n$ an integer number \cite{Skrinsky:1989ie}. This resonance condition can be observed by measuring  the cross section of $e^-e^-$ intra-beam scattering  (Touschek effect) with detectors located along the ring:  A sharp break occurs at $\omega_d$ due to the  larger cross section for polarized than for unpolarized electrons. The beam energy $E$ is calculated from $\omega_d$, using eq. (\ref{eq:Omega}).

Resonance depolarization was first applied at the Budker Institute of Nuclear Physics in Novosibirsk on the 625 MeV VEPP-2 storage ring in 1970 \cite{Blinov:2022nfg}, and became standard at other storage rings (CESR, DORIS, LEP). The most accurate measurement of the $J/\psi$ mass  of 2 ppm was achieved at the VEPP-4M storage ring with the KEDR detector  \cite{Anashin:2015rca}.

\section{The fifth quark}
\label{sec:bottomq}
\begin{figure}[htb]
\includegraphics[width=0.48\textwidth]{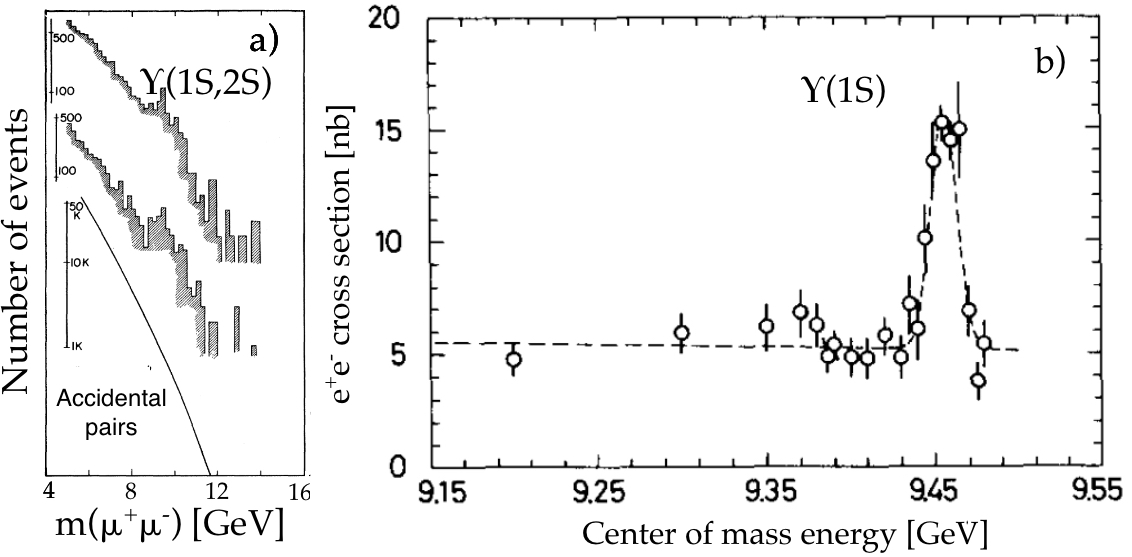}
\centering
\caption[]{a) The 10 GeV enhancement observed in $pA\to  \mu^+\mu^- X$ \cite{E288:1977xhf} with two settings of the spectrometer fields; b) The $\Upsilon(1S)$ reported by PLUTO in $e^+e^-$ collisions at DORIS \cite{Pluto:1978tuc}. The width of  8 MeV  is determined by the energy resolution of the storage ring.}
\label{DiscUps}
\end{figure}

The $b$ quark history started in 1977 at Fermilab  when an enhancement was observed  in the $\mu^+\mu^-$ mass spectrum with 400 GeV protons impinging on a Cu-Pt target  \cite{E288:1977xhf}. The muon momenta were analyzed by a double arm magnetic spectrometer equipped with meters long beryllium hadron filters to reduce multiple scattering. The structure around 10 GeV  (Fig. \ref{DiscUps}a)  was broader than the experimental resolution but could be fitted with two resonances around 9.4  and 10.2 GeV. This was the first evidence for two new heavy hadrons, the $\Upsilon(1S)$ and $\Upsilon(2S)$\footnote{An earlier measurement in the $e^+e^-$ mass spectrum reported a structure around 6 GeV \cite{Hom:1976cv} with a significance of about 3 standard deviations, but turned out to be due to a statistical fluctuation (today's common standard for claiming a discovery is 5 standard deviations).}. 

\begin{figure}[htb]
\includegraphics[width=0.48\textwidth]{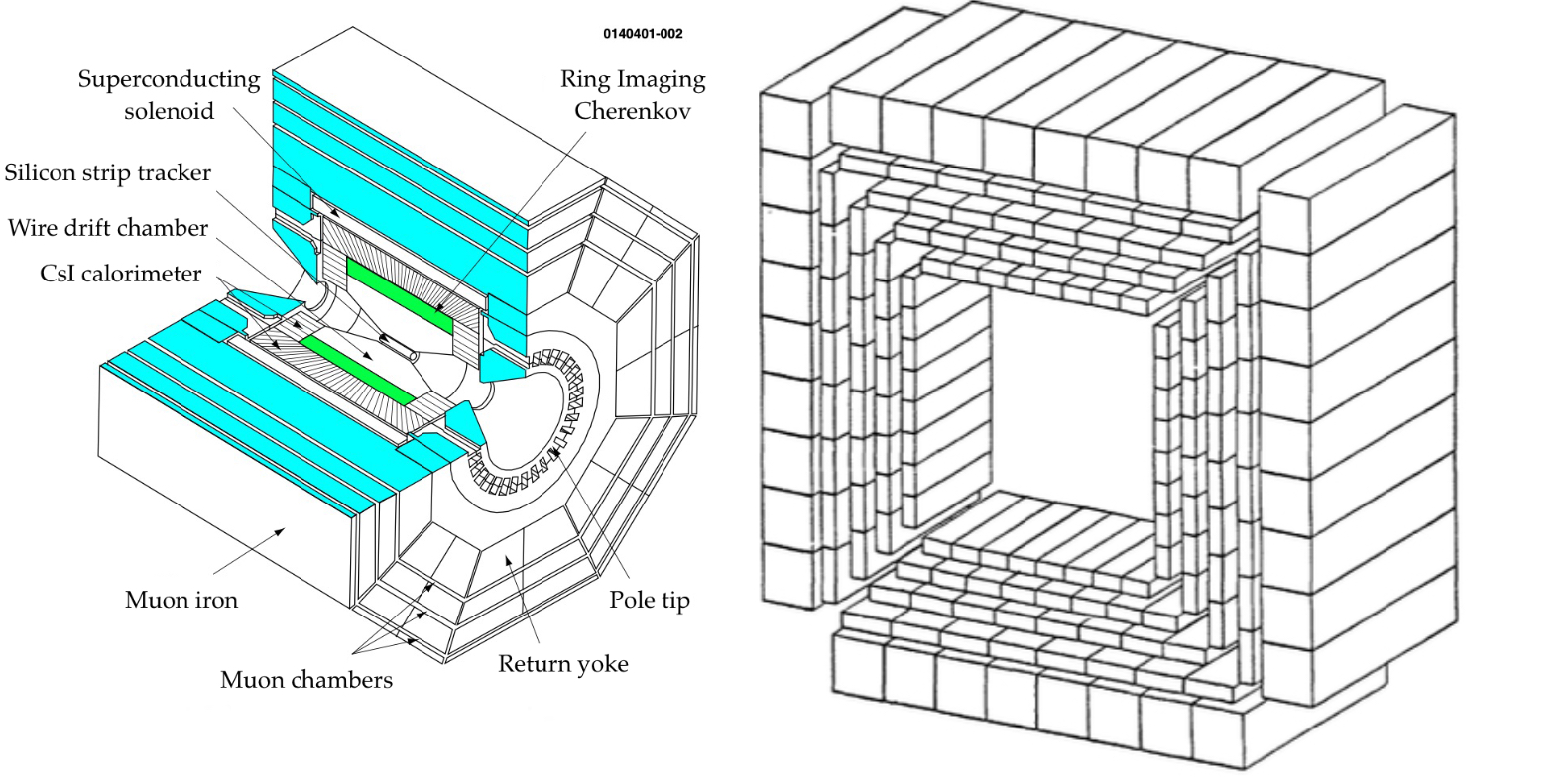}
\centering
\caption[]{Left: CLEO-III at CESR, a  magnetic solenoidal detector \cite{Briere2001} (Credit Cornell University). Right:  CUSB, a non-magnetic detector made of a large solid angle segmented array of NaI \cite{Bohringer:1980js}.
\label{CESRdetectors}}
\end{figure}

The $\Upsilon$  resonances listed in Table \ref{tab:psiups} were  subsequently observed  in $e^+e^-$ collisions at DORIS by the PLUTO \cite{Pluto:1978tuc}, DASP \cite{Darden:1978dk,Darden:1978ud} and  DESY-Heidelberg collaborations \cite{Bienlein:1978bg}, and at CESR by CLEO  \cite{CLEO:1980qvy} and CUSP \cite{Bohringer:1980js}. Fig. \ref{DiscUps}b shows the $\Upsilon(1S)$  in the $e^+e^-$ cross section by the super-conductive magnetic solenoid PLUTO.  The upgraded CLEO-III detector \cite{CLEO:1980tem} and the CUSB NaI array \cite{Finocchiaro:1980gy}  are shown in Fig. \ref{CESRdetectors} .

The abrupt  width increase  between $\Upsilon(3S)$ and $\Upsilon(4S)$ is a re\-petition of  the charm scenario described in section \ref{sec:charm}, due to the exis\-tence of a yet heavier quark, the bottom $b$ quark (sometimes called `beauty'): The three low mass $\Upsilon$'s in Table \ref{tab:psiups} are $b\bar{b}$ bound states  which cannot fragment into pairs of  `bottom' mesons $B$ ($\bar{b}q$)  and $\bar{B}$ ($b\bar{q}$), because they are below the production energy threshold which lies between the $\Upsilon(3S)$ and $\Upsilon(4S)$ masses. The electric charge of the $b$ quark is $-\frac{1}{3}e$, as can be deduced from the  ratio $R$ in Fig. \ref{Colored}. Also, according to  $\qqbar$ potential models the partial decay width of a vector meson into $e^+e^-$ is given by $\Gamma_{ee} = (16\pi\alpha^2Q^2/M^2)|\Psi(0)|^2$, where $M$ is the mass, $\Psi(0)$ the wave function at the origin and $Qe$ the electric charge. The measured $\Gamma_{ee}$ $\simeq$ 1.3 keV of the $\Upsilon(1S)$ is consistent with $|Q|=\frac{1}{3}$. 

The $\Upsilon(1S)$ mass was measured within $\pm$ 0.6 MeV at VEPP-4   \cite{Artamonov:1982mb} using resonance depolarization. 

\subsection{Bottom hadrons}
\begin{figure}[htb]
\includegraphics[width=0.25\textwidth]{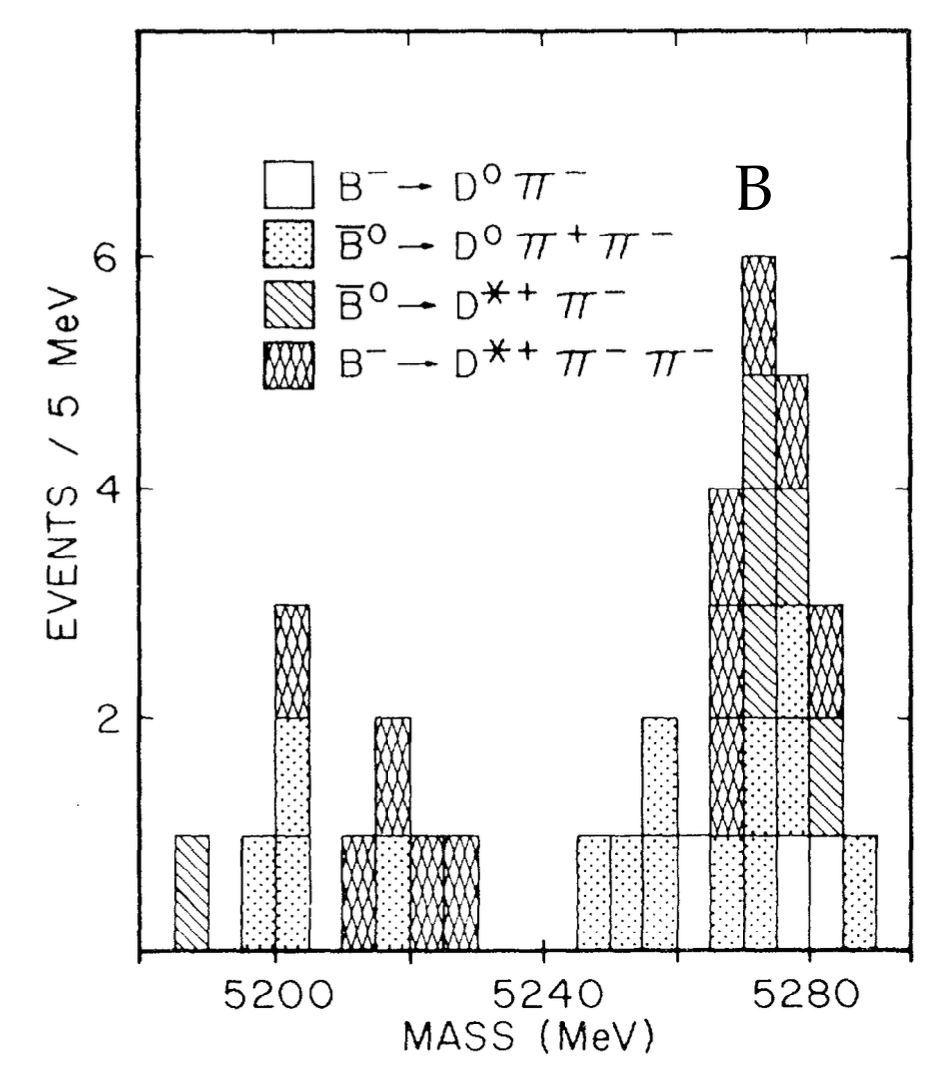}
\centering
\caption[]{Mass distribution of $D$ and $D^*$ plus one or two pions showing the $B$ signal  in  $e^+e^-\to\Upsilon(4S)$   around 5280 MeV (the charge conjugates $B^+$ and $B^0$ are implicitly assumed) \cite{CLEO:1983mma}. 
\label{Bmesons}}
\end{figure}

The $b$ quark decays dominantly to a charm quark ($b\to c W^+$) and therefore  $B$ mesons decay mainly to $D$ mesons. The $B$ meson was discovered by CLEO at the CESR $e^+e^-$ collider running on the $\Upsilon(4S)$ resonance,  searching for  $\Upsilon(4S)\to B\bar{B}$,  $B\to D$ or $D^*$, associated with one or two charged pions, and the $B$ carrying half the resonance energy \cite{CLEO:1983mma}. The mass distribution is shown  in Fig. \ref{Bmesons}.  

The $J^P=0^-$ $B(5280)$ mesons form the two isodoublets $\bar{B}^0 (b\dbar)$,  $ B^- (b\ubar)$  and $B^+  (u\bbar)$, $B^0 (d\bbar)$ and the two isosinglets
$\bar{B}^0_s (b\sbar)$, $B^0_s (s\bbar)$, $B^+_c (c\bbar)$ and  $B^-_c (b\cbar)$. The first evidence for the  $B_s(5367)^0$  was reported  by OPAL at LEP  in $Z^0$ decays \cite{OPAL:1992zsd}, while CDF at the TEVATRON measured its mass with the decay to $J/\psi \,\phi\to\mu^+\mu^- K^+K^-$ \cite{CDF:1993pzh}. The $B_c(6274)^\pm$ is difficult to observe due to its small production cross section.  The first two candidates were reported  by OPAL at LEP in $J/\psi\,\pi^+$ among  the debris of $Z^0$ decays into $b\bar{b}$ \cite{OPAL:1998gdf}. The $B$ mesons can be identified by their decay vertices in high energy experiments, thanks to their relatively long lifetimes ($\sim$1 ps).

The corresponding spin 1 mesons are named  $B^*$. The mass difference between $B^*$ and $B$ is only 45 MeV, hence the $B^*$ decays exclusively to $B\gamma$.  Several orbitals excitations of the $B$ mesons have been observed, as well as
baryons with one $b$ and two light   ($u$, $d$, $s$)  quarks \cite{ParticleDataGroup:2026cfk}.  The first  bottom baryon to be established was the $(udb)$ $\Lambda_b(5619)^0 \to J/\psi\Lambda$ by the UA1 Collaboration at the CERN $\bar{p}p$ collider \cite{UA1:1991vse}.   

The third quark family was expected from the existence of three lepton flavors. However, the partner of the $b$ -- the much heavier top quark $t$ -- decays too rapidly ($t\to W^+b$  within 10$^{-23}$s)  to build top hadrons, except possibly close to the $t\bar{t}$ threshold at 345 GeV.

\section{Quarkonium}
\label{sec:quarkonium}
\begin{figure}[htb]
\includegraphics[width=0.20\textwidth]{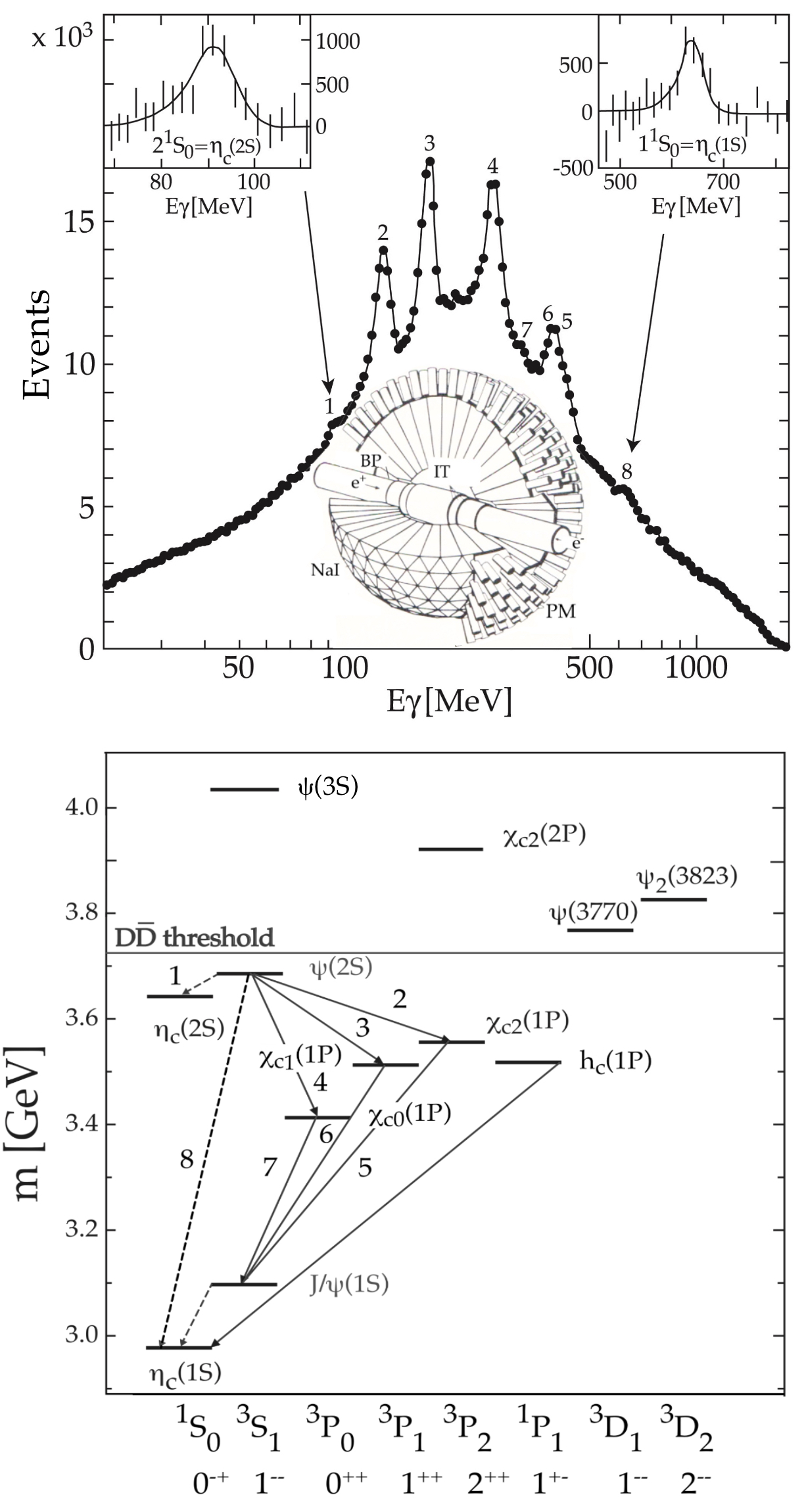}
\centering
\caption[]{Top: Radiative decay spectrum from the $\psi(2S)$ observed  in $e^+e^-$ collisions  \cite{KONIGSMANN1986243}. The inset shows the Crystal Ball NaI(Tl) assembly with beam pipe (BP), ionization tracker (IT) and photomultiplier tubes (PM). The $\gamma$ lines with widths determined by detector resolution correspond to the transitions in the bottom plot. Bottom: The $\ccbar$ spectrum as known in 2024.  Solid lines are E1 and dashed lines M1 $\gamma$-transitions. The quantum numbers are given in spectroscopic notation $^{2S+1}\ell_J$ and as $J^{PC}$. 
\label{Charmonium}}
\end{figure}

Table \ref{tab:psiups} lists the radial excitations of the charmonium ($c\bar{c}$) and bottomonium  ($b\bar{b}$) vector mesons with quark-antiquark angular momenta $\ell=0$ (S-states). Orbital excitations  ($P$, $D$, etc. with $\ell > 0$) have also been observed: For charmonium  the first indications stem from DORIS \cite{BRAUNSCHWEIG1975407}  and SPEAR \cite{Feldman:1975bq} searching for mono-energetic $\gamma$-transitions from the $\psi(2S)$ resonance in $e^+e^-$ collisions. Convincing evidence for the $\chi_{cJ}$ ($\ell = 1$) mesons with spins $J=0,1,2$  was obtained at SPEAR in 1977 with an array of NaI(Tl) crystals \cite{Biddick:1977sv}. Fig. \ref{Charmonium} (top) shows the $\gamma$ spectrum measured at the $\psi (2S)$ resonance by the Crystal Ball detector at SPEAR, a ball of 672 NaI crystals covering a solid angle of nearly 4$\pi$ installed at the $e^+e^-$ collision point  \cite{Partridge:1980vk}. The transition lines are observed above a large $\pi^0\to\gamma\gamma$ background. They constitute a direct proof for the existence of cons\-tituents inside hadrons.

The identification of the spin singlet $h_c(1P)$ was more difficult since radiative transitions $\psi(2S) \to \gamma h_c(1P)$ and $h_c(1P) \to \gamma J/\psi$ are forbidden by $C$-parity conservation, and the branching fraction for the isospin violating $\psi(2S) \to \pi^0 h_c(1P)$ is expected to be small. A handful of events associated with a $J/\psi$ and pions were observed in 1986 at the expected $h_c(1P)$ mass  in $\bar{p}p$ annihilations at the CERN ISR, which ran briefly with antiprotons in 1981--1983 \cite{R704:1986siy}. The narrow  $h_c(1P)(3525)$  decaying to  $\gamma\eta_c(1S)$ was finally established with 13 events in 2005 at the Fermilab $\bar{p}$ accumulator, where antiprotons collided with protons from a hydrogen jet target \cite{Andreotti:2005vu}.  The data set was later substantially increased in $e^+e^-$ collisions by CLEO-c \cite{CLEO:2008ero} and BES III
\cite{BESIII:2010gid,BESIII:2022tfo}, using the decay $\psi(2S)\to h_c(1P)\,\pi^0$ [$h_c(1P) \to \eta_c(1S)\,\gamma]$.

\begin{figure}[htb]
\includegraphics[width=0.48\textwidth]{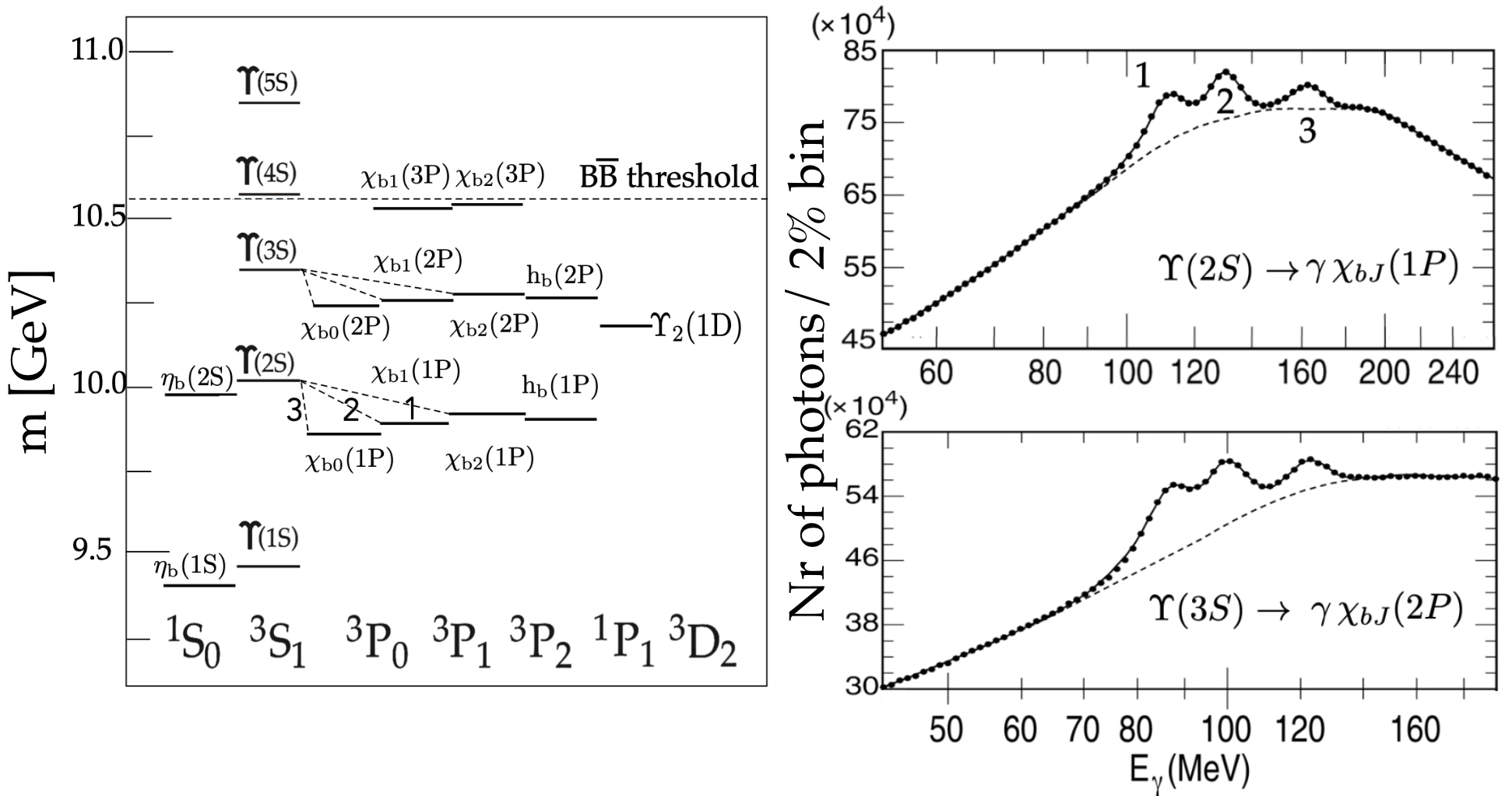}
\centering
\caption[]{Left: The known bottomonium states in 2024. 
Right: Radiative decay spectrum from the $\Upsilon(2S)$ and $\Upsilon(3S)$ observed  in $e^+e^-$ collisions  \cite{CLEO:2004jkt}. The $\gamma$ lines correspond to the transitions shown on the left. \label{Bottomonium}}
\end{figure}

In the bottomonium sector, the $\chi_{bJ}(2P)$ levels were the first to be reported at  CESR with the CUSP detector (shown in Fig. \ref{CESRdetectors})   in $\Upsilon(3S)$ radiative decay \cite{Han:1982zk}, followed by the observation of their transitions to the  $\Upsilon(1S)$ and $\Upsilon(2S)$ \cite{Eigen:1982zm}.   The radiative transitions $\Upsilon(2S)\to \gamma\chi_{bJ}(1P)$ and $\chi_{bJ}(1P)\to \gamma\Upsilon(1S)$ were published  shortly afterwards \cite{Klopfenstein:1983nx}. Fig. \ref{Bottomonium} shows the bottomonium spectrum and the inclusive high statistics photon spectrum measured by the upgraded CLEO-III experiment at CESR \cite{CLEO:2004jkt}, with  fits in the regions of the $1P$ and $2P$ $\chi_{bJ}$ states. The $\eta_b(1S)$, was established by BABAR at the SLAC PEP-II $e^+e^-$ collider  in inclusive radiative decay $\Upsilon(3S)\to \gamma\eta_b(1S)$ \cite{BaBar:2008dae}. 

\begin{figure}[htb]
\includegraphics[width=0.43\textwidth]{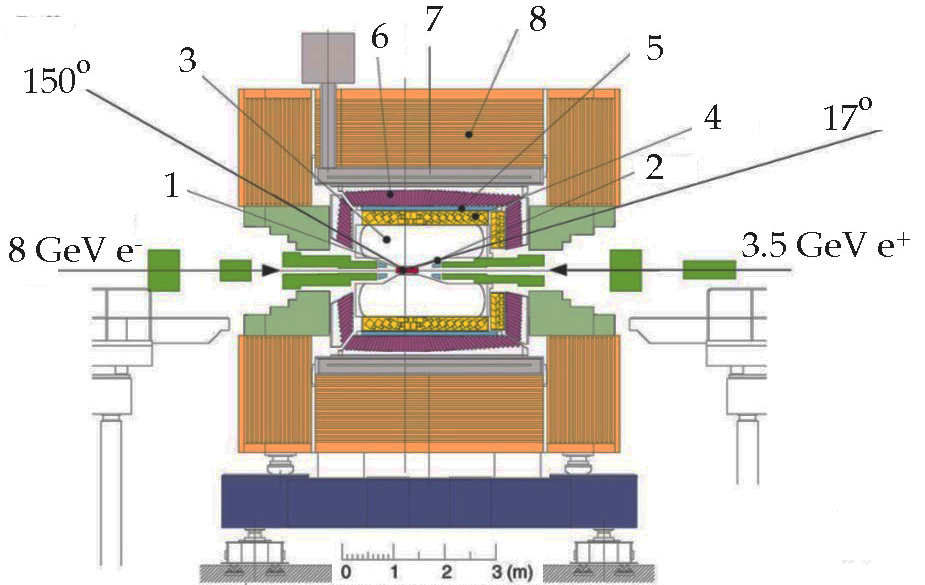}
\centering
\caption[]{The Belle detector  \cite{Belle:2000cnh}: 1--silicon vertex detector, 2--forward BGO calorimeter, 3--drift chamber, 4--particle identifier (aerogel), 5--time-of-flight counters, 6-CsI(Tl) $\gamma$-calorimeter, 7-superconducting solenoid, 8--muon detector (resistive plate counters).
\label{BelleDet}}
\end{figure}

\begin{figure}[htb]
\includegraphics[width=0.48\textwidth]{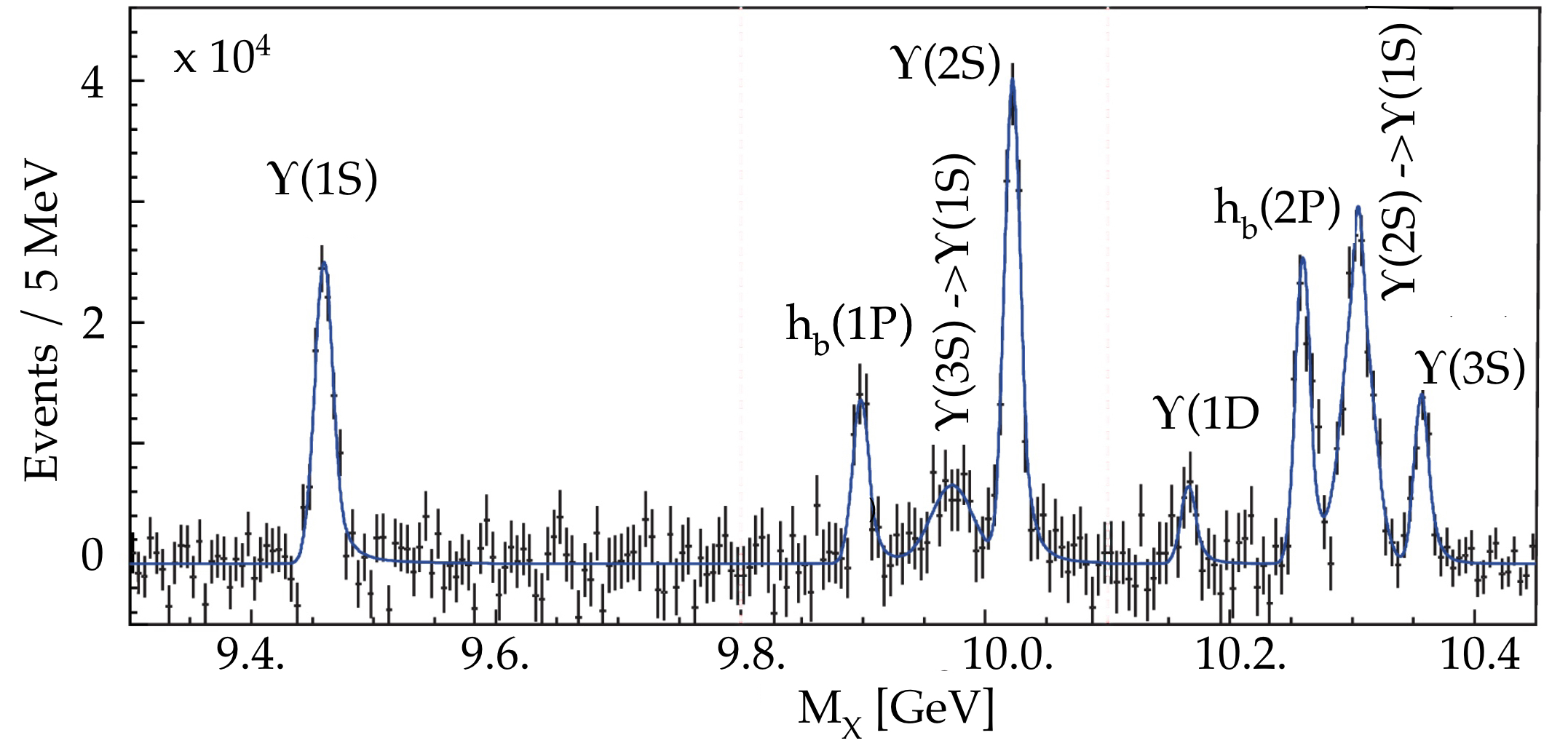}
\centering
\caption[]{The inclusive missing mass distribution in $\Upsilon(5S)\to \pi^+\pi^-\mathrm{M_X}$ after background subtraction \cite{Belle:2011wqq}.
\label{hb1P2P}}
\end{figure}

The $h_b(1P)$ and $h_b(2P)$ were first observed with the  Belle detector \cite{Belle:2011wqq} shown in Fig. \ref{BelleDet}. They were produced by exciting the $\Upsilon(5S)$ resonance and searching for missing masses recoiling against $\pi^+\pi^-$ pairs. The $\Upsilon(1S,2S,3S)$ are clearly visible in Fig. \ref{hb1P2P}, together with the $\Upsilon(5S)\to h_b(1P,2P)\,\pi^+\pi^-$ signals.  The $\eta_b(2S)$ was observed from the transition $h_b(2P)\to\gamma\eta_b(2S)$ by measuring the $h_b(2P)$ yield as a function of missing mass recoiling against $\pi^+\pi^-\gamma$ \cite{PhysRevLett.109.232002}.

Relative to masses the level spacings and the fine/hyperfine splittings are much larger for quarkonium than for hydrogen-like systems such as positronium (a bound  $e^+e^-$ pair). This is due to the binding force between quarks which is mediated by gluon exchange, while positronium is bound by the weaker photon exchange force. The mass difference (Table \ref{tab:psiups}) between the $2S$ and $1S$ levels for  $\ccbar$, $\bbbar$ (and even $\ssbar$) systems is around 600 MeV, hence  the attractive potential appears to be nearly flavor independent. Since the excitation energy is modest  compared to the $c$ and $b$ quark masses, quarkonium can be treated  by solving the Schr\"odinder equation with an appropriate potential model. 

\begin{figure}[htb]
\includegraphics[width=0.48\textwidth]{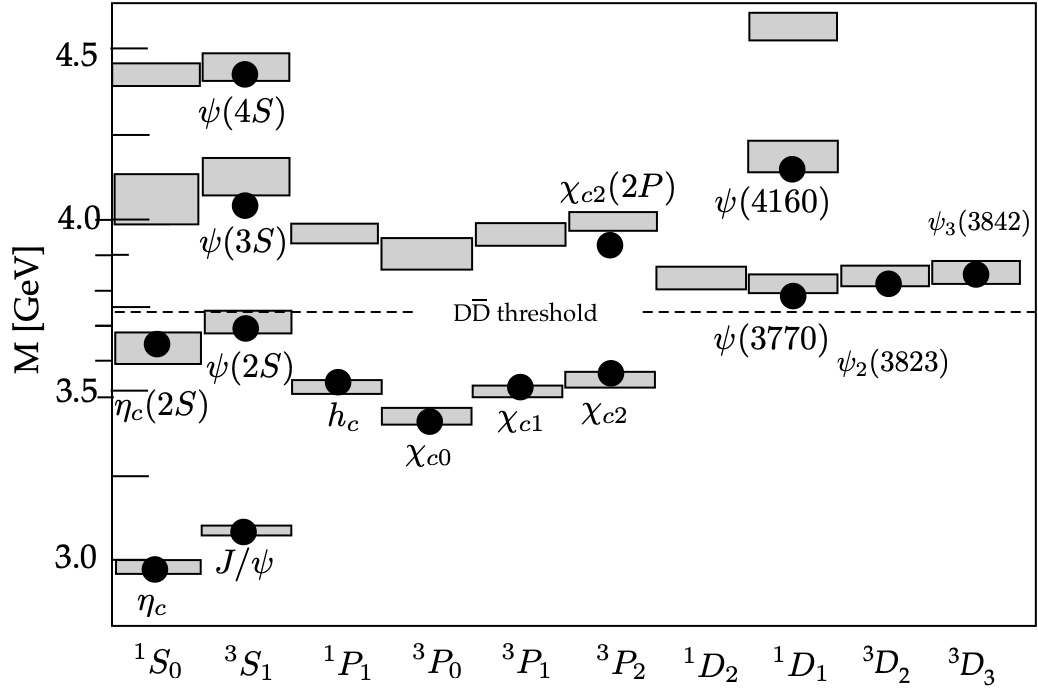}
\centering
\caption[]{The predicted charmonium mass spectrum (gray boxes) compared to the available data (blacks dots) (adapted from Ref.\cite{Pakhlova:2010zza}).
\label{qqbarpot}}
\end{figure}

For positronium (and hydrogen) the spacing of the energy le\-vels decreases as $1/n^2$, more rapidly with increasing  radial number $n$ than for quarkonium.  The  levels are nearly degenerate in positronium, while for quarkonium the $1P$ states lie   between the $1S$ and $2S$ levels (Fig. \ref{Charmonium}, \ref{Bottomonium}). Thus quarkonium  resembles  a 3-dimensional harmonic oscillator. An  ansatz for the potential energy of the $\qqbar$ pair is therefore between $ 1 / r $ (Coulomb-like) and $ r^2 $ (harmonic-like). The Cornell potential energy is of the form
\be
V(r) = -\frac{a}{r} + br ,
\label{eq:qqbarpot}
\ee
with $a$ = 0.10 GeV and $b$ = 0.93 GeV  when $r$ is expressed in fm \cite{Eichten:1979ms}. The $ 1 / r $ term  dominates at short distances and represents the exchange of a spin 1 gluon. The second term is responsible for the confining force from many gluons. For $\ell>0$  the fine splitting is due to the  interaction between angular momentum  and  quark spins ($LS$ coupling), and to the tensor interaction  between the  magnetic moments. The description of quarkonia by potentials of the form (\ref{eq:qqbarpot}) is impressive, as shown in Fig. \ref{qqbarpot} for charmonium.

The spin singlet $h_c(1P)$ is not affected by the $LS$ and tensor splittings and therefore its mass should coincide with the spin ave\-raged mass of the three $\chi_{cJ}(1P)$ states (3525.3 MeV). This is in impressive agreement with the 2024 best measured value of  the $h_c(1P)$ mass by the BESIII experiment at the Beijing BEPC $e^+e^-$ collider: 3525.32 $\pm$ 0.16 MeV \cite{BESIII:2022tfo}. The  spin averaged masses of the  $\chi_{bJ}(1P)$ and $\chi_{bJ}(2P)$ are also consistent with the $h_b(1P)$ and $h_b(2P)$ masses from Belle \cite{PhysRevLett.109.232002}.

\section{Exotic hadrons}
\label{sec:exotics}
QCD predicts  the existence of more complex structures than  $q\bar{q}$ pairs and quark trios: Glueballs are made exclusively of gluons bound by the exchange of color. Hybrid mesons  are $q\bar{q}$ pairs bound by an excited gluon ($q\bar{q}g$).  Several strong candidates have been reported. Although their decay patterns may be different, they are expected to mix with  $q\bar{q}$ hadrons lying in the same mass range and having the same quantum numbers, which makes the identification difficult. However, some may have `exotic' quantum numbers forbidden for $q\bar{q}$, {\it e.g.} $J^{PC}=1^{-+}$. Since 2015 candidates for pentaquark baryons ($qqqc\bar{c}$) have been reported by LHCb at CERN \cite{ParticleDataGroup:2026cfk}.

\subsection{Glueballs}
\begin{figure}[htb]
\includegraphics[width=0.48\textwidth]{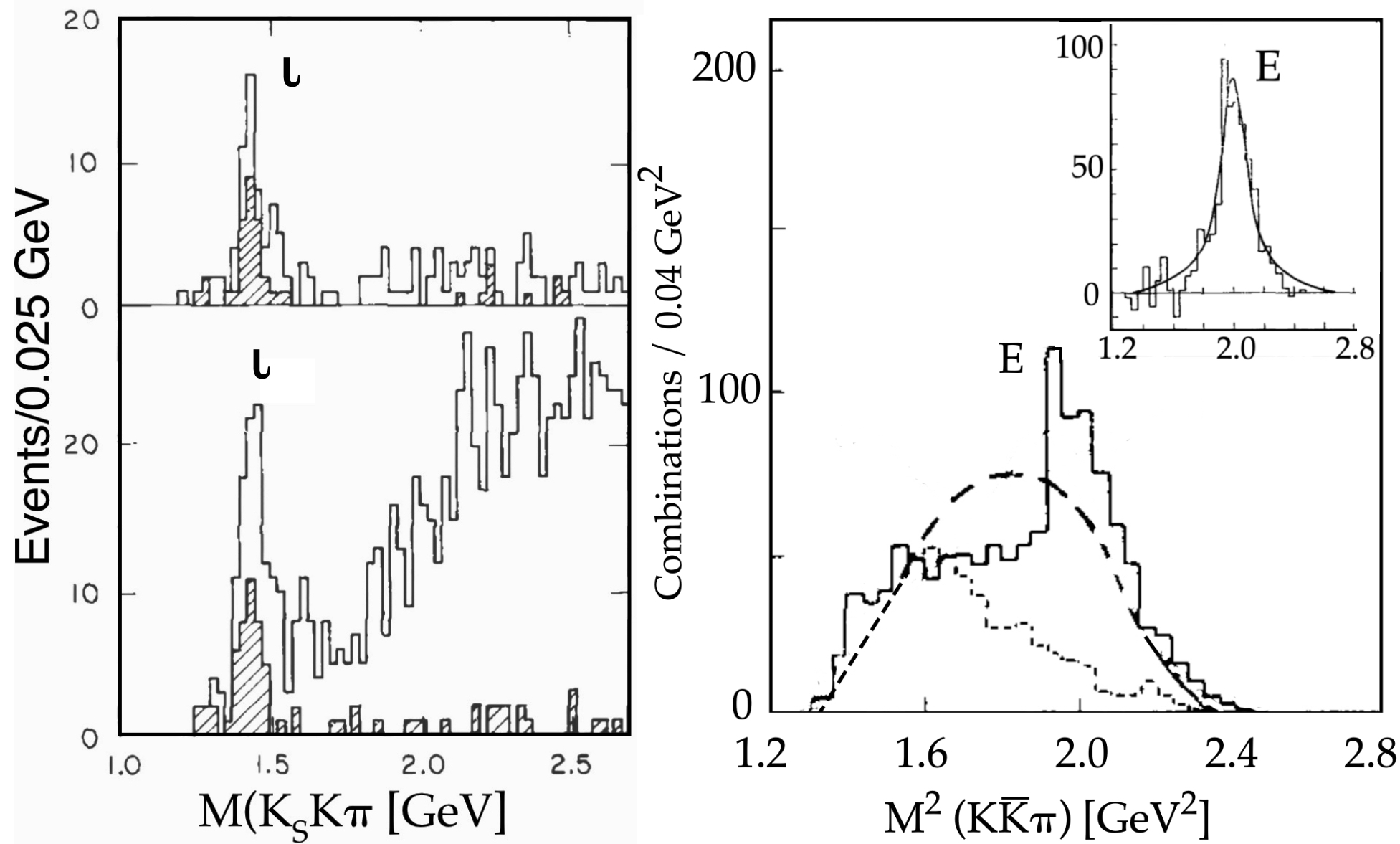}
\centering
\caption[]{Discovery of the $E/\iota$ meson. Left: The $K_SK^\pm\pi^\mp$ invariant mass in $J/\psi\to \gamma K_SK^\pm\pi^\mp$ with (top) and without (bottom) fitted $\gamma$ and when requiring a $K_SK^\pm$ mass below 1.05 GeV (shaded area) \cite{SCHARRE1980329}. Right: $K\bar{K}\pi$ squared mass distribution in $\bar{p}p\to K_SK^\pm K^\mp\pi^+\pi^-$. The dashed curve shows the phase space distribution. The histogram is for  $K_SK^\pm\pi^\mp$, the dotted one for $K_SK^\pm\pi^\pm$. The inset shows the difference and the fit (adapted from  Ref.\cite{Baillon:1967aa}). 
\label{Eiota}}
\end{figure}

The search for glueballs started in 1980 with  great excitement, following the observation by Mark II at SPEAR of a $0^{-+}$ meson around 1440 MeV in $J/\psi\to \gamma K\overline{K}\pi$, the  $\iota$ (iota) \cite{SCHARRE1980329}. The signal  (Fig. \ref{Eiota}, left) was believed to be a glueball  since radiative $J/\psi$ decay proceeds via the annihilation of the $c\bar{c}$ pair into gluons, which then hadronize into light quarks. A compatible state at 1425 MeV had already been observed earlier in $\bar{p}p$ annihilation at rest with the Saclay hydrogen BBC at the CERN PS \cite{Baillon:1967aa}. Fig. \ref{Eiota} (right) shows the then called $E$ meson decaying to $K_SK^\pm\pi^\mp$ in $\bar{p}p$ annihilation at rest into $K_SK^\pm K^\mp\pi^+\pi^-$. A $\gamma\rho$ enhancement at 1424 MeV in  $J/\psi\to\gamma\gamma\rho$ was also seen by  BESII  \cite{BES:2004pec}. 

At the epoch of the $E/\iota$ the MIT model predicted the $0^{-+}$ glueball to lie in the 1400 MeV region \cite{Jaffe:1976aa}. However, predictions from lattice gauge theories now  prevail: the $0^{-+}$ glueball is predicted at a much higher mass  of 2600 MeV. Instead, the ground state glueball is predicted to be  $0^{++}$ at about 1700 MeV, followed by a $2^{++}$ around 2400 MeV  \cite{Chen:2005mg}.

There are in fact two  isoscalar $0^{-+}$ mesons around 1440 MeV, the $\eta (1405)$ and $\eta (1475)$, the former  decaying mainly to $
a_{0}(980)\pi$ and the latter mainly to $K^{\ast }(892)\overline{K}$. Their natures remain elusive. Both states were confirmed by BESIII from a high statistics data sample in $J/\psi\to\gamma K_SK_S\pi^0$ \cite{BESIII:2022chl}. The $\eta(1405)$ decaying to $\eta\pi\pi$ was also observed by Crystal Barrel at LEAR \cite{CrystalBarrel:1995kfe}. 
There is room for only two isoscalar $0^{-+}$ radials, one  being the $\eta(1295)$ as the first radial excitation of the $\eta$ (Table \ref{tab:scalars}). The $\eta (1405)$ and $\eta (1475)$ compete for the second slot, the first radial excitation of the $\eta'$.  Are they perhaps the manifestation of one state only, induced by a kinematical (triangular) singularity \cite{Du:2019idk}?

\begin{table}[htb]
\begin{center}
\caption[]{The first radial excitations of the pseudoscalar mesons and the lowest lying $\ell=1$ scalar  mesons .}
\begin{tabular}{c c l l l}
\hline
 & & $I=1$ & $I=\frac{1}{2}$ & $I=0$ \\
 \hline
$0^{-+}$ & $2^1S_0$  & $\pi(1300)$ & $K(1460)$ & $\eta(1295, 1405, 1475)$ \\
$0^{++}$ &   & $a_0(980)$ & $K_0^*(700)$ & $f_0(500)$, $f_0(980)$ \\
$0^{++}$ & $1^3P_0$  & $a_0(1450)$ & $K_0^*(1430)$ & $f_0(1370,1500,1710)$ \\
\noalign{\vskip 1mm} 
\hline
\end{tabular}
\label{tab:scalars}
\end{center}
\end{table}

We now deal with the $0^{++}$ sector which involves the ground state glueball. The established $\ell = 1$ scalar mesons are shown in Table \ref{tab:scalars}. The ones below 1 GeV are often interpreted as tetraquarks ($qq\bar{q}\bar{q}$) or $K\bar{K}$ bound states, while  the $1^3P_0$ $\qqbar$ mesons lie above 1 GeV, where three isoscalar mesons compete for the two available slots. 
The isovector $a_0(980)$ (formerly known as $\delta$) was established in low energy $\bar{p}p$ annihilation into $\delta^\pm(\to K_SK^\pm)\pi^\mp$ with the Saclay hydrogen BBC at the CERN PS \cite{Astier:1967aa}.  
Its mass of about 1 GeV coincides  with the $K\bar{K}$  threshold and is not well determined. It also decays to $\eta\pi$, first spotted in $K^-n\to\Lambda\pi^-\eta$ with a kaon beam and a deuteron target \cite{Miller:1969aa}.  Its  isoscalar sibling $f_0(980)$ (formerly known as $S^*$), decaying to $K\bar{K}$ and $\pi\pi$, was observed in an amp\-litudes analysis of $\pi^+p\to\pi^+\pi^-({\rm or}\,  K^+K^-)\,\Delta^{++}$, using a 7.1 GeV/c pion beam \cite{Protopopescu:1973sh}. The $f_0(500)$ (aka $\sigma$) and the $K^*_0(700)$ (aka $\kappa$) are very broad $\pi\pi$, respectively $K\bar{K}$, resonances. Masses and widths have been measured  by many experiments, in particular  by upgraded versions of the BES detector. The width of  the $f_0(500)$ is $\sim$ 400 MeV from $J/\psi\to \gamma 3\pi$ \cite{BESIII:2016tdb} and that of the  $K^*_0(700)$  $\sim$ 600 MeV from $J/\psi\to K_SK_S\pi^+\pi^-$ \cite{BES:2010soq}.

\begin{figure}[h]
\includegraphics[width=0.48\textwidth]{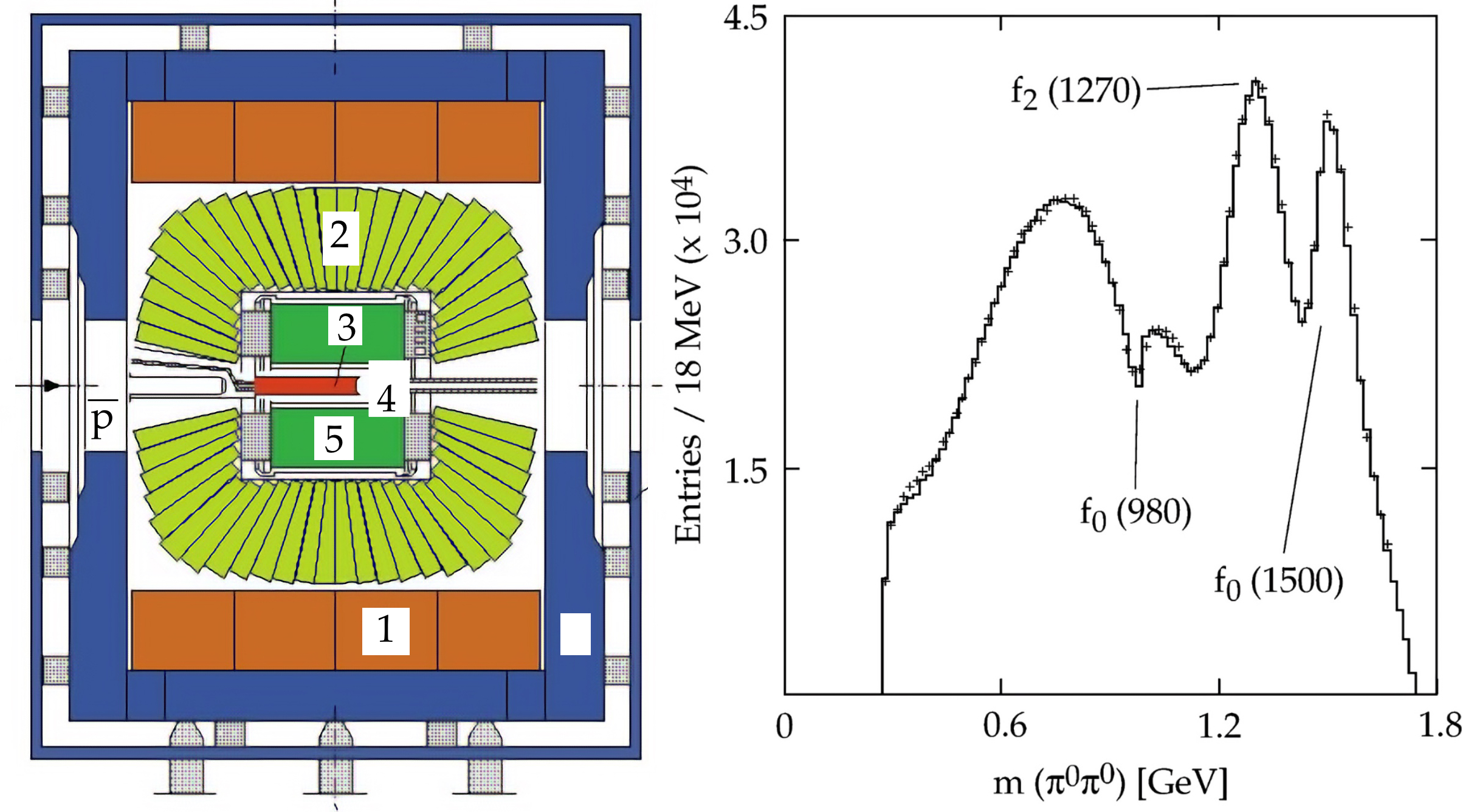}
\centering
\caption[]{Left: The Crystal Barrel detector. 1-coil, 2-CsI(Tl) 
barrel, 3-H$_2$ target, 4-proportional chambers, 5-drift chamber \cite{CrystalBarrel:1992qav}. Right: $\pi^0\pi^0$ mass distribution in  $\bar{p}p\to 3\pi^0$ at rest   showing  the $f_0(1500)$ in $\bar{p}p$ annihilation at rest into $3\pi^0$, together with the spin 2 $f_2(1270)$ and the $f_0(980)$ interfering destructively with the  broad $f_0(1370)$. The crosses are the data, the histogram is the fit \cite{Amsler:1995aa}.
\label{CBdetector}}
\end{figure}

The  $a_0(1450)$ and $f_0(1500)$ were discovered and the very broad  $f_0(1370)$ firmly established by Crystal Barrel at the Low Energy Antiproton Ring (LEAR) at CERN (Fig. \ref{CBdetector}, left)\footnote{A scalar structure  at 1590 MeV had been observed earlier by the GAMS collabo\-ration at CERN  in $\pi^-p\to \eta\eta n$ with 38 GeV pions   \cite{Serpukhov-Brussels-AnnecyLAPP:1983xdr}.} . A  pure conti\-nuous beam of 200 MeV/c antiprotons from LEAR (momentum bite $\Delta p/p \sim 10^{-3}$) entered a  conventional 1.5 T solenoidal magnet and stopped in a liquid hydrogen target. Fig. \ref{CBdetector} (right) shows the $2\pi^0$ invariant mass distribution in $\bar{p}p$ annihilation  into $3\pi^0\to 6\gamma$. A simultaneous fit was performed to $\bar{p}p\to 3\pi^0, 2\pi^0\eta$ and $2\eta\pi^0$, which demanded the three scalar resonances $a_0(1450)\to\eta\pi^0$, $f_0(1370)$ and $f_0(1500)$ \cite{CrystalBarrel:1995dzq}. For a review of Crystal Barrel data see Ref.\cite{Amsler:1997up}.

\begin{figure}[htb]
\includegraphics[width=0.48\textwidth]{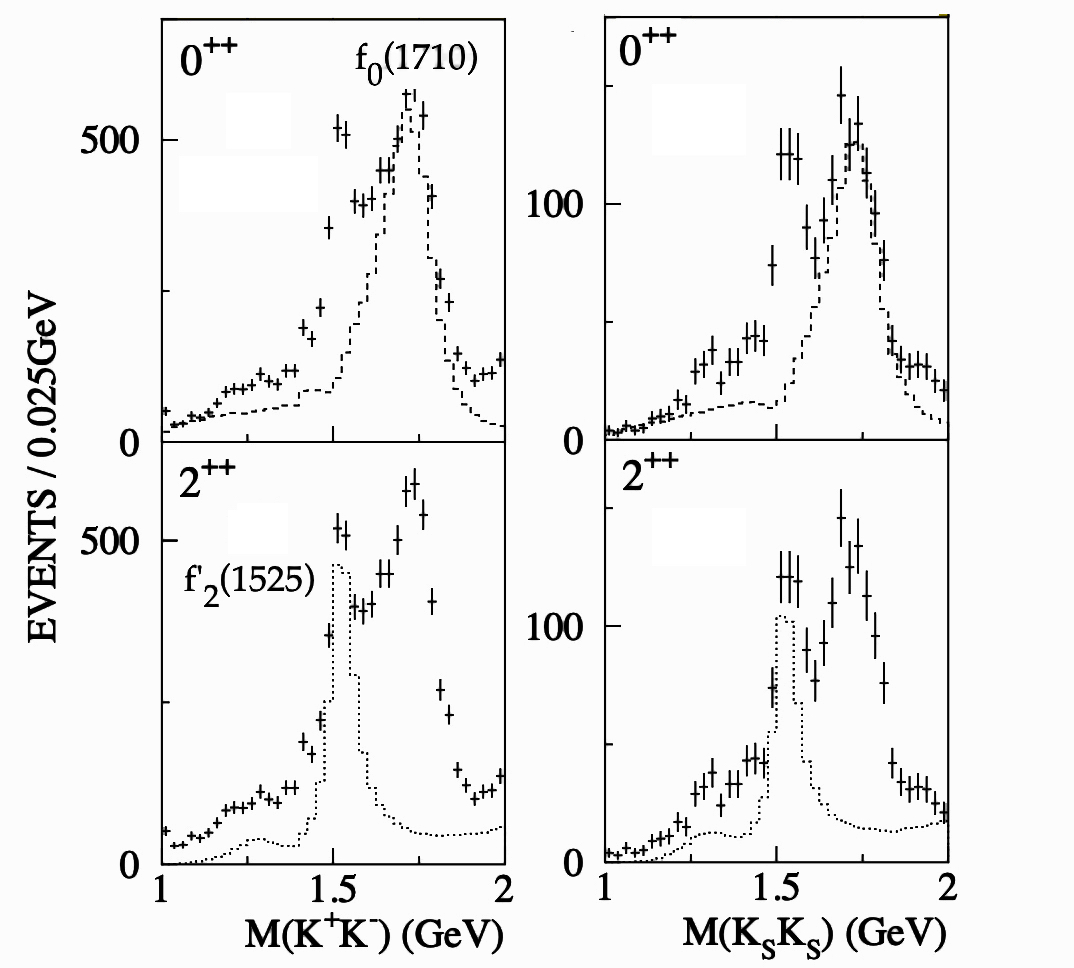}
\centering
\caption[]{$K\bar{K}$ mass distributions in $J/\psi\to\gamma K\bar{K}$ from BES. The dashed distributions show the fitted contributions to the amplitudes for a spin 0 $f_0(1710)$ and a spin 2  $f'_2(1525)$ \cite {BES:2003iac}.
\label{KKbarBES}}
\end{figure}

\begin{figure}[htb]
\includegraphics[width=0.40\textwidth]{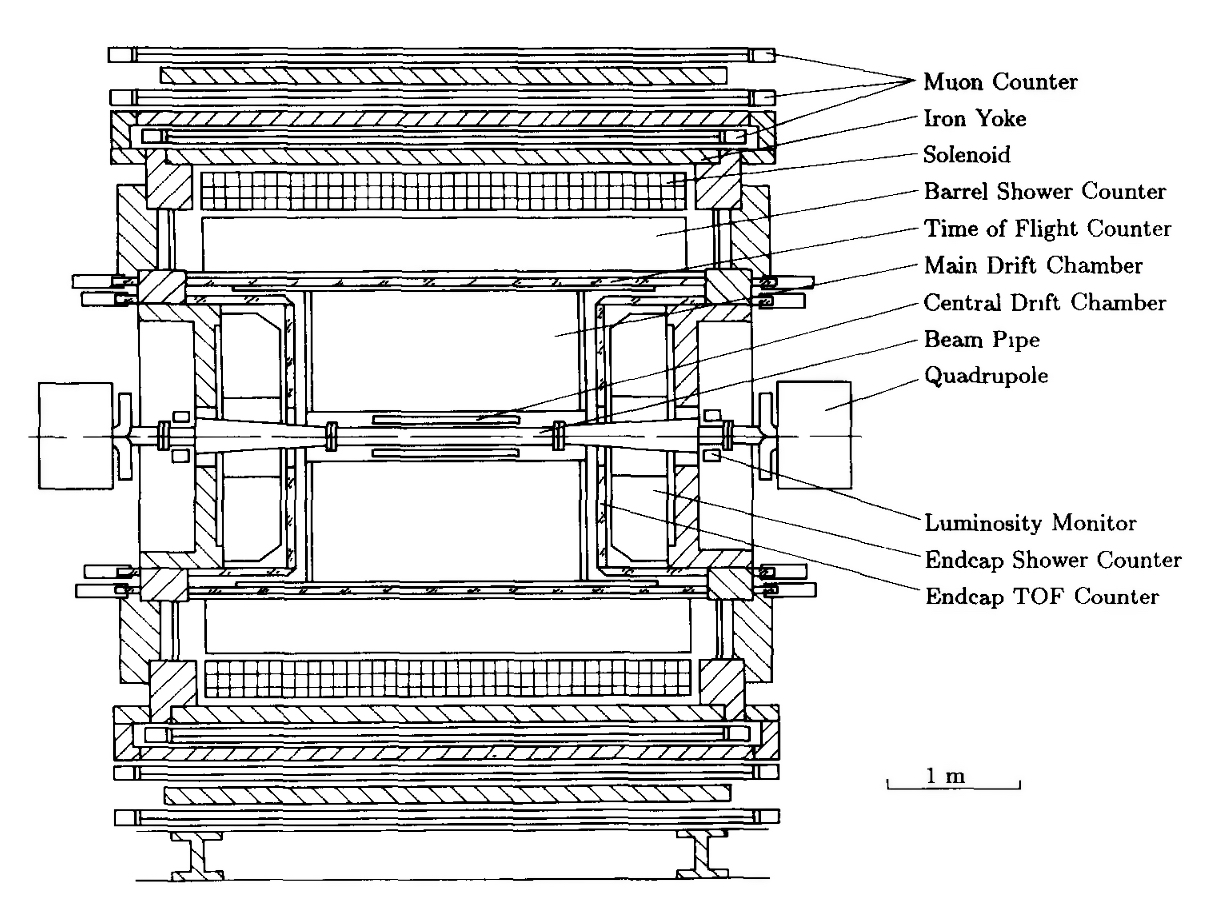}
\centering
\caption[]{The BES detector at the  BEPC $e^+e^-$ collider \cite{BES:1994bjo}. 
\label{BESdetector}}
\end{figure}

The $f_0(1710)$ has a long history of controversial spin 0 or 2 assignment. The issue was solved in favor of $J = 0$ from a partial wave  analysis of the centrally produced $K^+K^-$ system in $pp$ interactions at 300 GeV, measured with the Omega Spectrometer at the CERN SPS \cite{French:1999tm}, reinforced by an analysis of radiative $J/\psi$ decay into $K^+K^-$ and $K_SK_S$ at BES \cite {BES:2003iac}. The $K\bar{K}$ mass spectra in Fig. \ref{KKbarBES} prove that the signal around 1700 MeV is indeed spin 0, while spin 2
is  assigned to the established $f'_2(1525)$ tensor meson, known to decay dominantly to $K\bar{K}$. The BES detector is shown in Fig. \ref{BESdetector}.  For a review of BES data see Ref.\cite{Ye:2020aa}.

Which of the two scalars, $f_0(1500)$ or $f_0(1710)$, then qualifies to be the $0^{++}$ glueball?  This issue is not settled yet: The gluon rich $J/\psi$ radiative decay into $\pi^+\pi^-$ and $\pi^0\pi^0$  is more prominent for  the $f_0(1710)$  \cite{Ablikim:2006db}, which is then favored over the $f_0(1500)$. On the other hand, the strong $K\bar{K}$ signal of the $f_0(1710)$ is indicative of a large $\ssbar$ content. 
However, the three mesons overlap and mix: The $f_0(1370)$ and $f_0(1710)$ have been proposed as $\bar{q}q$ states mixed with glue, while the $f_0(1500)$ is dominantly glue mixing with $q\bar{q}$ \cite{Amsler:1995td,Close:2001ga}.  
When central production data from WA102 (CERN) are combined with hadronic $J/\psi$ decay data from BES, glue is  shared between the $f_0(1370)$ (mainly $u\bar{u}+d\bar{d}$),  $f_0(1500)$ (mainly glue) and $f_0(1710)$ (dominantly $s\bar{s}$) \cite{Close:2005vf}.

\subsection{Light exotic mesons}
\label{hybrids}
Two  isovectors with exotic quantum numbers $1^{-+}$ have been reported by several experiments, the $\pi_1(1400)$  and the $\pi_1(1600)$.   The $\pi_1(1400$ was observed in the $\eta\pi^0$ $P$-wave by the GAMS collaboration at CERN with 100 GeV pions in $\pi^-p\to\eta\pi^0 n$ \cite{Alde:1988aa}.  A reso\-nating
$1^{-+}$ $\eta\pi$ $P$-wave was also required by Crystal Barrel in $\overline{p}p\to\pi^0\pi^0\eta$ \cite{CrystalBarrel:1999reg}.  

The  $\pi_1(1600)$ decaying to $\rho\pi$ was reported by the B852 collaboration at BNL with 18 GeV pions 
in $\pi^-p\to\pi^-\pi^-\pi^+p$ \cite{E852:1998mbq},  later confirmed by COMPASS  at CERN with 190 GeV pions  \cite{COMPASS:2009xrl}, and in $\pi^-p\to\eta'\pi^-p$ \cite{2015303}.   
A coupled channel analysis of the $\eta\pi$  and $\eta'\pi$ data  led later to a single broad $\pi_1$ state  at 1564 $\pm$ 89 MeV, with a width of 492 $\pm$ 115 MeV \cite{JPAC:2018zyd}.  A 188 MeV broad  $1^{-+}$ enhancement  at 1855 MeV in $J/\psi$ radiative decay into the  $\eta\eta'$ P-wave has been reported by BESIII \cite{BESIII:2022iwi} which  could be  the isoscalar partner of the $\pi_1(1600)$. 

Lattice calculations predict that the lightest hybrid with the exo\-tic quantum numbers $1^{-+}$  lies around 1.9 GeV (see {\it e.g.} Ref.\cite{MILC:1997usn}). For a review on hybrid mesons we refer to Ref.\cite{Meyer:2015eta}.

\begin{figure}[htb]
\includegraphics[width=0.48\textwidth]{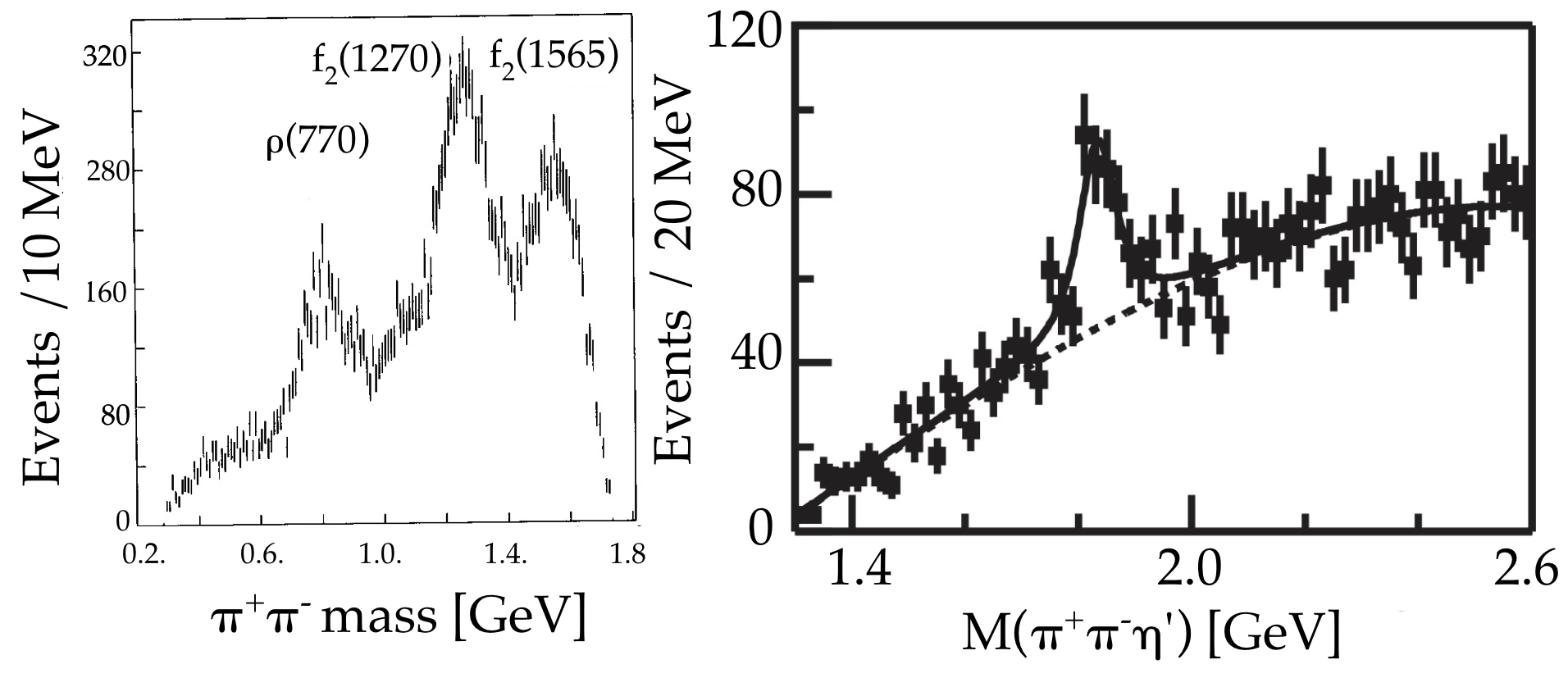}
\centering
\caption[]{Left: The $f_2(1565)$ in $\bar{p}p\to\pi^+\pi^-\pi^0$ annihilation at rest in hydrogen gas, associated with $L$ x-rays to the $2p$ atomic levels (to select annihilation with one unit of angular momentum) \cite{MAY1989450}. Right: The pseudoscalar state observed by BES in $J/\psi\to\gamma\pi^+\pi^-\eta’$ at 1835 MeV \cite{PhysRevLett.95.262001}. 
\label{AX1835}}
\end{figure}

Nucleon-antinucleon  bound states have been predicted, based on the attractive ($N\bar{N}$) meson exchange \cite{BUCK197947}.  The $f_2(1565)$, a supernumerary meson in the tensor nonet, has been discovered by the ASTERIX experiment at LEAR in $\bar{p}p$ annihilation at rest into three pions (Fig. \ref{AX1835}, left).  This meson has also been seen by the OBELIX experiment at LEAR in $\bar{p}p$ \cite{BERTIN1997476} and $\bar{n}p$ annihilation \cite{PhysRevD.57.55} into three pions, and is a 
good candidate for such a state  \cite{BUCK197947}. A $0^{-+}$ candidate has been reported by BES around 1835 MeV in  $J/\psi\to \gamma \bar{p}p$ \cite{PhysRevLett.91.022001} and $J/\psi\to\gamma\pi^+\pi^-\eta’$ \cite{PhysRevLett.95.262001} (Fig. \ref{AX1835}, right) which also qualifies as a bound  $\bar{p}p$ candidate.

\subsection{Heavy exotic mesons}
The early 2000s saw the emergence of mesons containing both light $\qqbar$ and heavy $Q\bar{Q}$ ($\ccbar$ or $\bbbar$) pairs. They could be compact $\qqbar Q\bar{Q}$ states (tetraquarks), $q\bar{Q} -\bar{q}Q$ hadron-hadron loosely bound states or hadro-charmonia  ($\ccbar$ pairs surrounded by clouds of lighter $\qqbar$ pairs). They were initially called $X,Y$ or $Z$, but a new naming scheme was introduced in 2024 \cite{ParticleDataGroup:2026cfk} to cope with the plethora of candidates. We shall mention here some of the historical ones which have been fully established, see Ref.\cite{ParticleDataGroup:2026cfk} and the reviews therein on non-$q\bar{q}$ and $Q\bar{Q}$ mesons and pentaquark baryons.   

\begin{figure}[htb]
\includegraphics[width=0.48\textwidth]{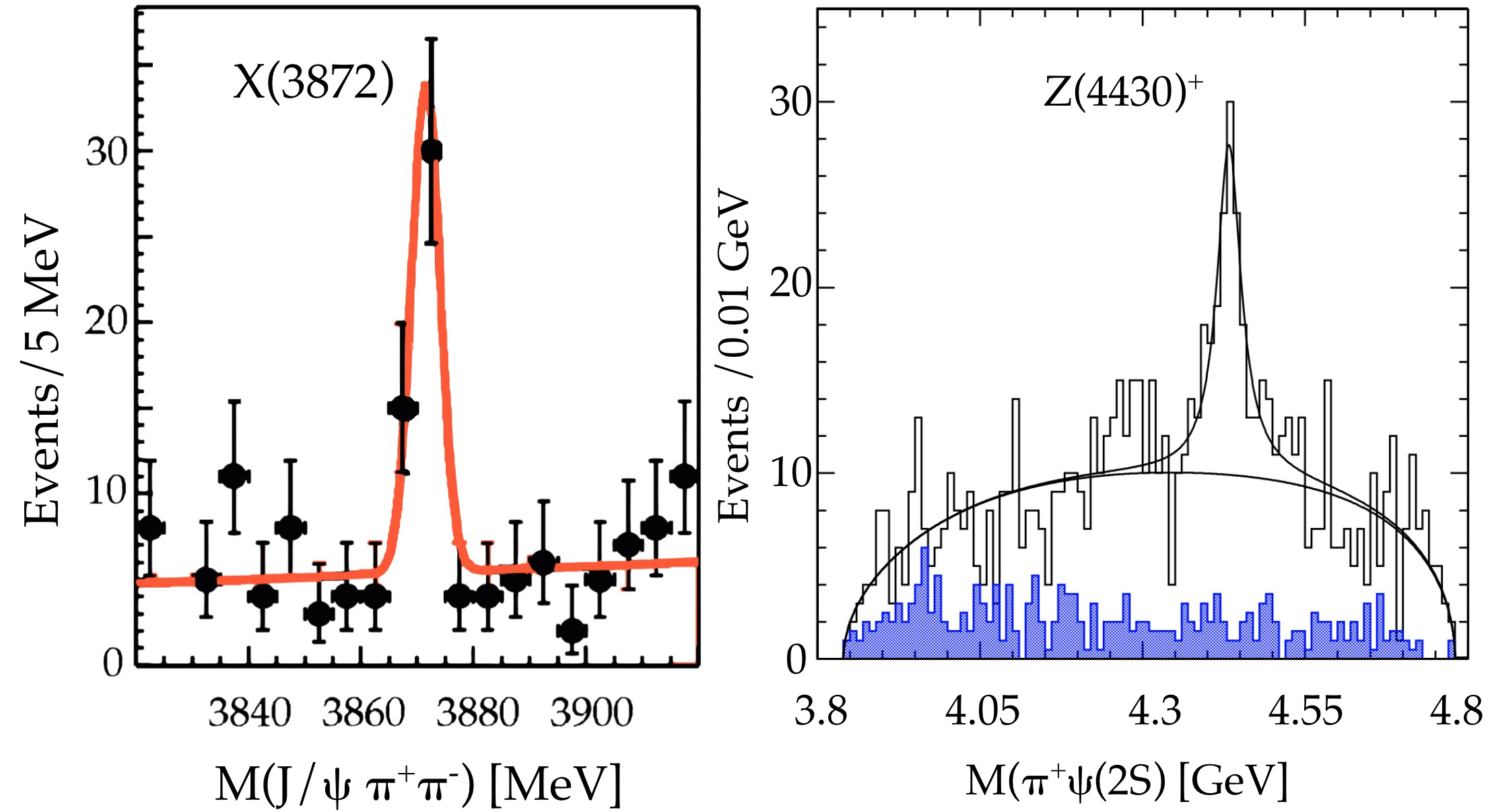}
\centering
\caption[]{Left: Observation of the $X(3872)\to J/\psi\,\pi^+\pi^-$ by Belle in $B^\pm\to K^\pm J/\psi\,\pi^+\pi^-$\cite{Belle:2003nnu}; Right: Observation of the $Z(4430)^+\to\pi^+\psi(2S)$ \cite{Belle:2007hrb}. \label{Tetraheavy}}
\end{figure}

The exotic revolution  started effectively in 2003 with the discovery of the $\chi_{c1}(3872)$ (aka $X(3872)$)  by Belle in $B^\pm$ decay to $K^\pm J/\psi\,\pi^+\pi^-$ (Fig.  \ref{Tetraheavy}, left). The $B$ mesons were produced in $e^+e^-$ collisions at the $\Upsilon(4S)$. It was confirmed by BABAR and by  experiments at CERN and Fermilab\footnote{A tentative evidence had been  reported earlier at Fermilab from high energy $\pi$-nucleus data \cite{E705:1993pry}.}, and its 
$1^{++}$ quantum numbers  were firmly established by LHCb \cite{LHCb:2015jfc}. The $X(3872)$ is  too light by $\sim$100 MeV to be the $2^3P_1$  charmonium state. On the other hand,  it is very narrow (1.2 MeV)  and lies at the $D^0\overline{D}^{*0}$  threshold, which suggests a $D^0\overline{D}^{*0}$ `molecule',  weakly bound by one pion exchange, as predicted many years ago  \cite{Tornqvist:1991ks}.

Mesons decaying into charmonia and  a charged pion cannot be $c\cbar$  states, but also contain charged  $q\bar{q}$ pairs. Fig. \ref{Tetraheavy} (right) shows the first observation of the $T_{c\bar{c}1}(4430)^+$ (aka $Z(4430)^+$) by Belle, containing both $c\cbar$ and charged $u\bar{d}$ pairs, and decaying to $\pi^+\psi(2S)$. This $J^P=1^+$ state is produced in $B\to K\pi^+\psi(2S)$.   The $T_{c\bar{c}1}(3900)^+$ (aka $Z_c(3900)^+$),  reported by BESIII \cite{BESIII:2013ris} and Belle \cite{Belle:2013yex}, is another prominent $1^+$ charged charmonium-like state produced in $e^+e^-\to J/\psi\,\pi^+\pi^-$ and decaying to $\pi^\pm J/\psi$.

Mesons that do not fit in the  $\bbbar$ spectrum have been discovered by Belle, the $T_{b\bar{b}1}(10610)^\pm$ (aka $Z_b(10610)^\pm$) and the $T_{b\bar{b}1}(10650)^\pm$ (aka $Z_b(10650)^\pm$)\cite{Belle:2011aa}. They were produced by running on the $\Upsilon(5S)$ resonance,  $\Upsilon(5S)\to Z_b(10610)^\pm\pi^\mp{\rm \ or\ } Z_b(10650)^\pm\pi^\mp$. Both mesons then decay to $\Upsilon(1S), \Upsilon(2S)$ or  $\Upsilon(3S)$ by emitting another charged pion. The quantum numbers $J^P = 1^+$ are favored from the angular distribution of the pion. As charged mesons they are not bottomonium states.  They  lie close to the $B^*\overline{B}$ and $B^*\overline{B}^*$ thresholds, respectively, and thus could be hadronic `molecules'.

\begin{figure}[htb]
\includegraphics[width=0.48\textwidth]{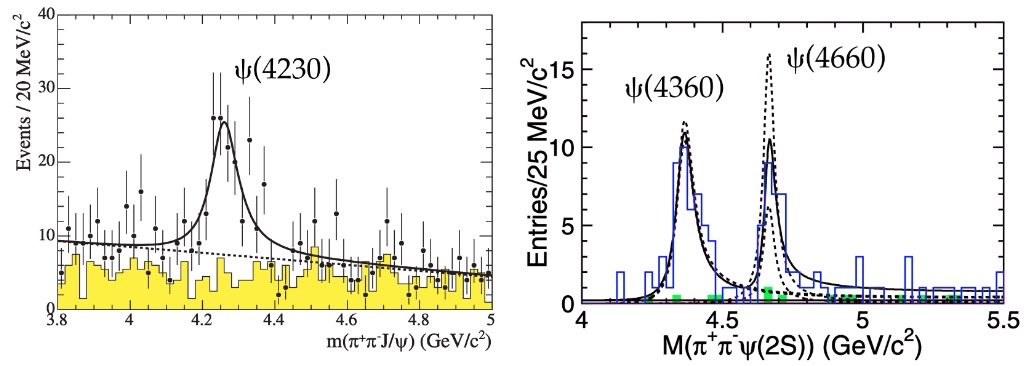}
\centering
\caption[]{Left: Mass distribution in $e^+e^-\to\gamma_{ISR}\,J/\psi\,\pi^+\pi^-$ ($J/\psi\to e^+e^-$ or $\mu^+\mu^-$) from BABAR \cite{BaBar:2005hhc}. The yellow histogram shows the contribution from the $e^+e^-$ or $\mu^+\mu^-$ side bands of the $J/\psi$. Right: Mass distribution in $e^+e^-\to\gamma_{ISR}\,\psi(2S)\,\pi^+\pi^-$ ($\psi(2S)\to J/\psi\,\pi^+\pi^-$) from Belle \cite{Belle:2007umv}.
\label{ISRpsi}}
\end{figure}

Isoscalar $1^{--}$ resonances have also  been found that do not decay into  $D\bar{D}$ pairs, albeit lying above the $D\bar{D}$ threshold: the $\psi(4230)$, $\psi(4360)$ and $\psi(4660)$ were discovered by searching for decays into $J/\psi$ or the $\psi(2S)$. The $\psi(4230)$ was first observed by BABAR by scanning the $e^+e^-$ cross section associated with the emission of a photon (initial state radiation, ISR), and looking for resonances decaying into $\pi^+\pi^-J/\psi$ (Fig. \ref{ISRpsi}, left).   Fig. \ref{ISRpsi} (right) shows the $\psi(4360)$  and the $ \psi(4660)$  from Belle, decaying to $\pi^+\pi^-\psi(2S)$. 

\section{Conclusions}
\label{sec:conclusions}
More than 400 mesons and baryons have been reported since the discovery of the pion and kaon in the 1940s. The evidence for quarks and gluons, albeit confined in hadrons, is overwhelming. Most hadrons are consistent with the traditional `naive’ $q\bar{q}$ and $qqq$ quark model involving five quarks.  However, exotic candidates  predicted by QCD have been found but their clear  identification as glueballs and hyb\-rids is still open. Since the observation of the X(3872) in 2003, one also knows that more complex structures such as tetraquarks and pentaquarks exist, which are open for further experimental and theoretical investigations.
For an excellent record on important experiments see Ref.\cite{Cahn:1989by} and the annual Review of Particle Physics \cite{ParticleDataGroup:2026cfk}. For details on the quark model and the phenomenology of hadron spectroscopy, see for example Refs.\cite{FrankClose, Amsler:2018zkm}.

\bibliography{reference}

\end{document}